\documentclass[12pt,letterpaper]{article}
\usepackage[margin=1in]{geometry}
\usepackage[onehalfspacing]{setspace}
\usepackage{graphicx}%
\usepackage{multirow}%
\usepackage{amsmath,amssymb,amsfonts, bm}%
\usepackage{amsthm}%
\usepackage{mathrsfs}%
\usepackage[title]{appendix}%
\usepackage{xcolor}%
\usepackage{textcomp}%
\usepackage{manyfoot}%
\usepackage{booktabs}%
\usepackage{algorithm}%
\usepackage{algorithmicx}%
\usepackage{algpseudocode}%
\usepackage{listings}%
\usepackage{array}
\usepackage{hyperref}
\usepackage[draft]{changes}
\usepackage{authblk}

\title{A probabilistic framework for crystal structure denoising, phase classification, and order parameters}

\author[1,2]{Hyuna Kwon\thanks{Email: \texttt{hkwon7@binghamton.edu}}}
\author[1]{Babak Sadigh}
\author[1]{Sebastien Hamel}
\author[1]{Vincenzo Lordi}
\author[1]{John Klepeis}
\author[1]{Fei Zhou\thanks{Email: \texttt{zhou6@llnl.gov}}}

\affil[1]{\small Lawrence Livermore National Laboratory, Livermore, CA, USA}
\affil[2]{\small Binghamton University (SUNY), Binghamton, NY, USA}

\begin{document}
\maketitle

\begin{abstract}
Atomistic simulations generate large volumes of noisy structural data, yet extracting phase labels and continuous order parameters (OPs) in a robust and general manner remains challenging. Existing tools are often specialized to a limited set of prototypes and split thermal-noise removal, phase classification, and OP construction into separate steps.
Here we present a unified probabilistic framework for analyzing noisy atomic configurations with respect to known crystal prototypes. The model predicts per-atom, per-prototype logits and aggregates them into a scalar log-probability (logP) landscape over atomic coordinates. Its gradient defines a conservative denoising field, while the logits provide local phase labels, prototype-resolved OPs, and ambiguity measures through logit margins. We train on AFLOW-mapped crystalline structures from the Materials Project with synthetic positional and elastic perturbations, then test extrapolation to stronger noise, finite-temperature disorder, point defects, water-ice coexistence, binary polymorphs, and shock-compressed Ti.
A single differentiable scalar model recovers prototype identity after denoising, tracks smooth transformations such as Bain and Burgers paths, and exposes low-confidence regions near defects and phase boundaries. This provides an integrated and extensible tool for analyzing complex atomistic simulations.
\end{abstract}

\section*{Introduction}
Atomistic simulations generate increasingly large and complex datasets for studying phase transitions, defect formation, and microstructural evolution~\cite{belonoshko2006melting, zepeda2017probing, shibuta2017heterogeneity, Zepeda-Ruiz2017N}. Advances in first-principles calculations, machine-learning interatomic potentials (MLIPs), and high-performance computing now enable routine simulations over long timescales. However, extracting physical insight from these datasets remains challenging. In particular, (i) assigning crystalline phase labels to individual atoms and (ii) defining continuous order parameters (OPs) that quantify structural order are difficult tasks in realistic configurations containing thermal disorder, defects, interfaces, and phase coexistence.

Significant progress has been made on crystal structure classification for ideal or weakly perturbed unit cells. The Curtarolo group, for example, has curated the AFLOW Encyclopedia of structural prototypes~\cite{Mehl2017CMS-prototypeP1,Hicks2018CMS-prototypeP2,Hicks2021CMS-prototypeP3,Eckert2024CMS-prototypeP4} and developed tools such as XtalFinder~\cite{Hicks2021nCM-XtalFinder}, which efficiently match relaxed primitive cells to known prototypes. For large-scale atomistic configurations, a range of local structural descriptors is widely used, including common neighbor analysis (CNA)~\cite{Honeycutt1987JPC-CNA}, bond-orientational OPs~\cite{Steinhardt1983PRB,Lechner2008JCP}, centrosymmetry analysis~\cite{Kelchner1998PRB-CSA}, and polyhedral template matching (PTM)~\cite{larsen2016robust}. While effective for a limited set of well-studied lattices such as BCC, FCC, and HCP, these approaches rely on hand-crafted heuristics and often degrade under strong thermal disorder, defects, or phase coexistence, leading to ambiguous or incorrect classifications~\cite{Hsu2024nCM-denoiser}.

Continuous OPs provide complementary scalar measures of structural order. Classical examples, such as Steinhardt-type bond-OPs, are widely used for characterizing liquid--solid transitions~\cite{Steinhardt1983PRB,Lechner2008JCP}. However, unlike the systematic cataloging for crystal structures, OPs are typically constructed on a case-by-case basis for specific systems, limiting their general applicability and interpretability in complex, heterogeneous datasets. This motivates a unified framework that can both identify crystal structures and provide continuous, phase-resolved measures of structural similarity.

Machine learning (ML) offers a promising route toward more general structure characterization. Prior work combined symmetry-invariant descriptors (e.g., SOAP, bispectrum) with neural networks to classify crystal structures or detect phase transitions~\cite{Ziletti2018NC,Geiger2013JCP,Defever2019CS,Fulford2019JCIM,Kim2020PCCP,Swanson2020SM,Doi2020JCP,Becker2021PRE,Leitherer2021NC,Chung2022PRM,Hernandes2022JPCM,Chapman2022nCM,Chapman2023NC,Aroboto2023APL,Moradzadeh2023JPCC}. Bayesian deep-learning approaches such as ARISE have further improved robust crystal-structure recognition by classifying a large number of structural classes while providing uncertainty estimates under noise and defects~\cite{Leitherer2021NC}. More recently, in our previous works~\cite{Hsu2024nCM-denoiser,Sun2024JCIM-ice}, we adapted the score-based diffusion models~\cite{Sohl-Dickstein2015-DPM,Ho2020-DDPM,Song2019NeurIPS-Generative}, enabling the removal of thermal noise by learning a denoising score field~\cite{Hsu2024nCM-denoiser,Sun2024JCIM-ice}.

Despite these advances, several limitations remain for general-purpose structure analysis. Currently, denoising and classification are treated as separate tasks, where a learned denoiser first refines atomic positions and a downstream classifier, which is trained separately if not readily available, assigns phase labels, potentially discarding subtle structural information \cite{Hsu2024nCM-denoiser,Sun2024JCIM-ice,Zaidi2023ICLR,New2024-crystaldenoising,Shen2024-denoising-pre-training}. In addition, most approaches focus on discrete classification, with limited or no continuous confidence measures to capture ambiguity near phase boundaries or in highly disordered regions. Finally, many models are specialized to specific systems or small sets of phases, limiting their broader applicability.

Energy-based and score-based models provide a natural statistical framework for integrating these tasks~\cite{LeCun2006-Energy,Grathwohl2020ICLR,Li2023-DiffusionClassifier,Chen2023-RobustClassifier} by defining a scalar ``energy'' function $E(\boldsymbol{r})$ over configurations whose gradient $\nabla_{\boldsymbol{r}} \log \hat{P}_\theta(\boldsymbol{r})$ drives denoising while its outputs encode structural preferences. This suggests that a single model can simultaneously assign phase probabilities, provide denoising directions, and define continuous OPs, avoiding the need for separate pipelines. The probabilistic formulation used here follows established energy-based modeling approaches; the novelty lies in its application to atomistic structure analysis and in the resulting unification of denoising, classification, and OP extraction.

In this work, we introduce a probabilistic framework for atomistic structure analysis based on a learned log-probability ($\log P$) over atomic configurations. By modeling the gradient of $\log P$ with respect to atomic positions, the approach naturally yields a conservative vector field that enables structure denoising, while the $\log P$ itself provides a basis for classification and continuous structural characterization. This formulation allows denoising, classification, and OP extraction to be performed within a single model, rather than as separate stages. We emphasize that our goal is not to achieve state-of-the-art classification accuracy; specialized classifiers may outperform our model on purely discriminative tasks. Instead, the focus is on unifying multiple structure analysis tasks within a single probabilistic framework. In terms of denoising performance, the present conservative log-probability model achieves accuracy comparable to a direct non-conservative score-based denoiser\cite{Hsu2024nCM-denoiser} trained with the same architecture and data, as confirmed by matched RMSE benchmarks. The advantage of the $\log P$ formulation is not more accurate denoising. It is that denoising is obtained from a conservative score field while phase classification and continuous logit-based OPs are obtained from the same scalar log-probability landscape.

The approach predicts per-atom, per-phase logits $l_{ac}$, where $a$ indexes atoms and $c$ indexes candidate crystal phases. Aggregating these across atoms via a log-sum-exp yields a total machine-learned log-density $\log\!\hat{P}_\theta(\boldsymbol{r})$, whose gradient defines a conservative score field $\boldsymbol{s}(\boldsymbol{r}) = \nabla_{\boldsymbol{r}}\log\!\hat{P}_\theta(\boldsymbol{r})$ for denoising, while the per-phase logits serve as physically motivated OPs measuring similarity to each class $c$. Phase labels are obtained directly by selecting the class $c$ with the largest $l_{ac}$, and ambiguous regions can be identified through low or mixed $l_{ac}$ values. Training follows the paradigm of MLIPs, combining a denoising score-matching loss~\cite{Vincent2011scorematching} (analogous to force matching) and a cross-entropy classification loss on the logits $l_{ac}$ (analogous to fitting per-structure energy labels in MLIP training). In contrast, our previous denoiser model directly predicts a non-conservative score field $\hat{\boldsymbol{s}}_\theta(\boldsymbol{r})$, similar to direct-force predictions of some force fields, with no explicit conservative log-probability structure~\cite{Hsu2024nCM-denoiser,Sun2024JCIM-ice}.

We implement this framework using a graph neural network (GNN) architecture based on MACE~\cite{Batatia2022mace,Batatia2023foundation}, trained on a curated subset of Materials Project structures~\cite{jain2013commentary} mapped to AFLOW prototypes~\cite{hicks2021aflow}. The model is trained using a combination of score-matching and classification objectives, allowing it to learn both structural refinement and phase discrimination within a single representation. We demonstrate that this approach achieves robust denoising and accurate classification across a range of crystalline systems, including thermally perturbed structures, interfaces, and phase transformations such as ice polymorphs and shock-compressed Ti. Beyond discrete labeling, the model provides continuous, defect-sensitive OPs that track structural evolution, offering a practical tool for analyzing noisy atomistic configurations. Throughout this work, we use $\boldsymbol{R}$ to denote discrete atomic configurations and $\boldsymbol{r}$ for continuous coordinates.

\begin{figure}
    \centering
    \includegraphics[width=1\linewidth]{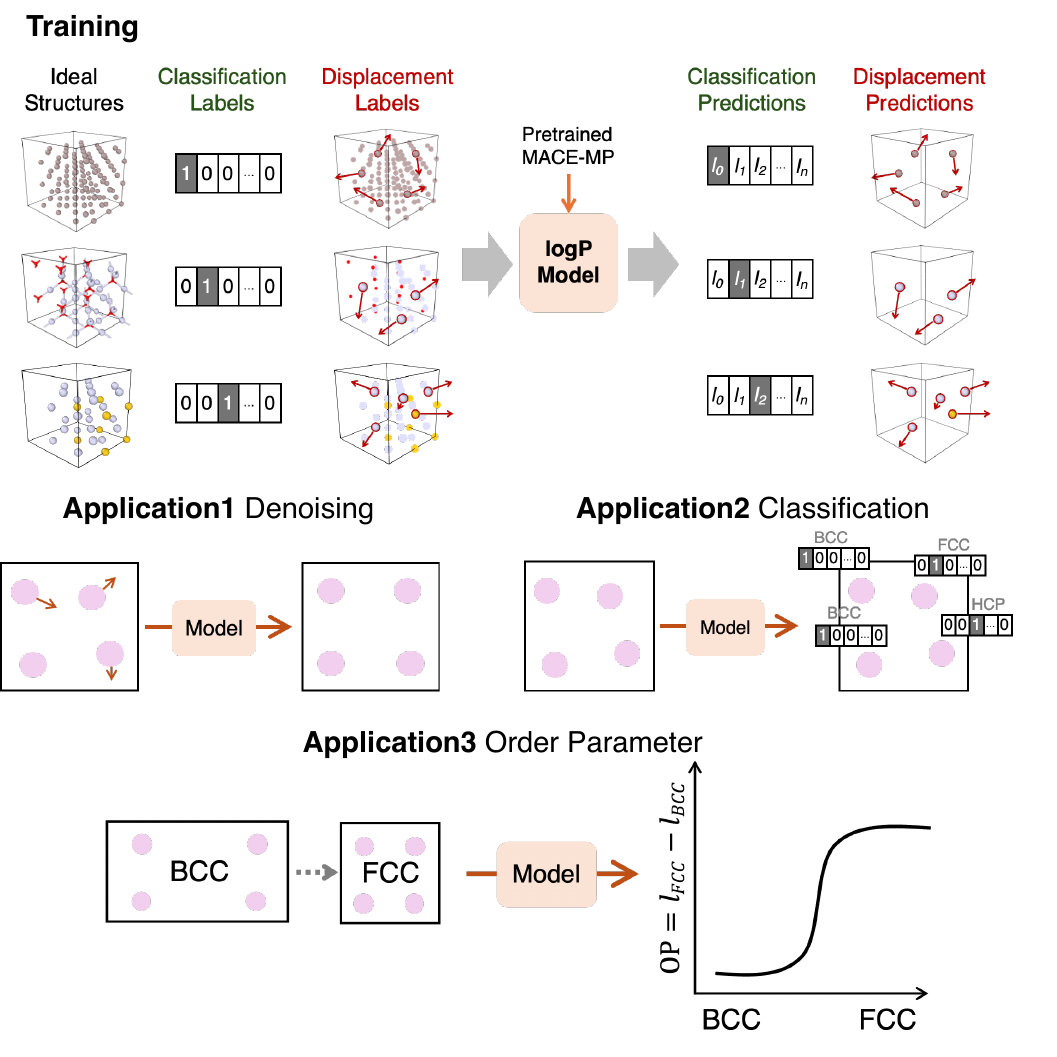}
    \caption{Overview of the log-probability ($\log \hat{P}_\theta$) model. Training uses ideal crystalline structures mapped to AFLOW prototypes, with two coupled objectives: (i) predicting per-atom, per-phase logits $l_{ac}$ guided by crystal class labels, and (ii) learning the conservative score field $\nabla_{\boldsymbol{r}}\log\!\hat{P}_\theta(\boldsymbol{r})$ of the aggregated log-density $\log \hat{P}_\theta$ from randomly displaced structures. At inference time, the same model can be used to iteratively denoise noisy configurations, assign phase labels from $\arg\max_c l_{ac}$, and evaluate per-atom $l_{ac}$ fields as continuous, phase-resolved OPs.}
    \label{fig:workflow}
\end{figure}

\section*{Methods}

\subsection*{Log-probability model.}
The $\log \hat{P}_\theta$ model predicts per-atom, per-phase logits (unnormalized scores prior to normalization) $l_{ac}$, from which we construct
a global log-density
\begin{equation}\label{eq:logP}
    \log \hat{P}_\theta(\boldsymbol{r}) = \sum_a \log \sum_c \exp\left(l_{ac}(\boldsymbol{r})\right).
\end{equation}
The per-phase logit $l_{ac}$ depends on the neighboring atoms and acts as continuous, phase-resolved OPs that quantify how similar each atom $a$ is to prototype $c$.
Throughout, we refer to $l_{ac}$ as (raw) logits and reserve ``log-probability'' for the aggregated
quantity $\log \hat{P}_\theta$.
This architecture, which yields
$C$ (number of classes) scalars per atom, is
analogous to per-atom decomposed energy in an MLIP, allows us to leverage established MLIP training pipelines based on derivative (score or force) matching. Additionally it ensures the model remains extensive and the resulting score field remains conservative. Placing the log-sum-exp inside the atomic sum allows different local environments in one configuration to favor different prototypes, which is essential for interfaces and phase coexistence. Eq.~\ref{eq:logP} should be interpreted as a local structural mixture. 
The gradient
\begin{equation} \label{eq:score}
    \boldsymbol{s}_\theta (\boldsymbol{r}) = \nabla_{\boldsymbol{r}} \log \hat{P}_\theta(\boldsymbol{r})
\end{equation}
defines a conservative score field used for denoising.

In practice, we instantiate this model using the MACE equivariant graph neural network architecture. For pretrained models, the MACE-MP checkpoint is used to initialize the embedding, message-passing, and readout layers, and the scalar energy readout is modified to produce $C$ per-atom outputs corresponding to the prototype logits. The resulting decoder $\hat{D}_\theta$ predicts
\begin{align}
{l}_{ac} = \hat{D}_{\theta,ac}(\boldsymbol{z}_a),
\end{align}
where $\boldsymbol{z}_a$ denotes the equivariant latent representation of atom $a$. All model parameters, including the MACE representation layers and the newly added prototype-resolved decoder, are then optimized jointly under the combined score-matching and classification objective. For models trained from scratch, the same architecture is initialized randomly and trained with the same objective. The computational cost of inference is comparable to the original MACE model, with only a small additional overhead from the prototype-resolved output channels. More details on the ML model are provided in the Supporting Information.

\subsection*{Dataset and augmentations.}

To ensure consistency and relevance in structural representations, we curated a subset of the Materials Project~\cite{jain2013commentary} dataset by filtering entries to match crystal prototypes from the AFLOW Encyclopedia~\cite{hicks2021aflow}. Specifically, we include only Materials Project entries that (i) can be mapped to an AFLOW prototype and (ii) lie within 0.1~eV/atom above the convex hull, thereby focusing on experimentally plausible or metastable phases. The AFLOW Encyclopedia includes only prototypes observed in at least ten experimentally or computationally verified compounds, so this filtering step removes rare, idiosyncratic structures (e.g., CsMg\textsubscript{149}) that hinder generalization, and yields a dataset enriched in structurally meaningful crystalline motifs.

To account for physically realistic variations in lattice parameters, we applied a small random elastic deformation combining isotropic scaling and symmetric strain. For each structure, we first sampled an isotropic scale factor $s \sim \mathcal{U}[0.9,\,1.1]$ and then drew a strain tensor
\begin{align}
    E_{ij} \sim \mathcal{U}(-\delta_{\text{strain}},\,\delta_{\text{strain}}),
\end{align}
with $\delta_{\text{strain}} = 0.05$. To avoid introducing spurious rigid-body rotations, we symmetrized the strain tensor as
\begin{align}
    E \leftarrow \tfrac{1}{2}\,(E + E^\mathsf{T}),
\end{align}
and formed the total deformation gradient
\begin{align}
    T = s\,(I + E).
\end{align}
The deformation $T$ was applied consistently to both the cell vectors and Cartesian atomic positions, followed by periodic wrapping of atoms back into the simulation cell. This augmentation exposes the model to moderate volumetric and shear strains while preserving the underlying prototype symmetry and periodicity.

Unless otherwise noted, we use $\sigma_{\max} = 0.15$~\AA\ as the maximum positional noise
scale when constructing noisy configurations for score matching (see below). For each
primitive cell, we build an approximately cubic supercell containing $\sim$210 atoms to
provide sufficient local environments for graph-based learning.

\subsection*{Training objectives and optimization.}

The total training loss combines a score-matching term and a classification term,
\begin{align}
    \mathcal{L}
    &= \mathcal{L}_\text{sm} + w_\text{cl}\,\mathcal{L}_\text{cl},
\end{align}
where $\mathcal{L}_\text{cl}$ encourages the logits $l_{ac}$ to match the known prototype
label $c$ of the ideal structure $\boldsymbol{R}_0$, and $\mathcal{L}_\text{sm}$ trains a fixed-scale restoring field toward the ideal structure.
During training, we construct noisy configurations by adding Gaussian noise to the ideal
structure $\boldsymbol{R}_0 = \{\boldsymbol{r}_{0a}\}_{a=1}^N$: 

\begin{align*}
   \tilde{\boldsymbol{R}}  = \boldsymbol{R}_0 +  \Delta \boldsymbol{r} = \boldsymbol{R}_0 + \sigma_n \boldsymbol{\epsilon},  \  \sigma_n \sim \mathcal{U}(0.001, \sigma_{\max}), \ 
    \boldsymbol{\epsilon}   \sim \mathcal{N}(0, \mathbf{I}),
\end{align*}
where $\sigma_n$ is sampled per structure to represent various noise scales at different temperature.

For classification, we use a per-atom cross-entropy loss,
\begin{equation}
    \mathcal{L}_\text{cl}
    = -\,\mathbb{E}_{\boldsymbol{R}_0, c, \sigma_n, \boldsymbol{\epsilon}}
      \left[
        \frac{1}{N}
        \sum_{a=1}^N
        \log \frac{\exp(l_{ac})}{\sum_{c'} \exp(l_{ac'})}
      \right].
\end{equation}
where $c$ is the AFLOW prototype label associated with $\boldsymbol{R}_0$.

The score-matching loss is
\begin{equation}
    \mathcal{L}_\text{sm}
    = \mathbb{E}_{\boldsymbol{R}_0,\sigma_n,\boldsymbol{\epsilon}}
    \left[
        \frac{1}{N}
        \sum_{a=1}^N
        \left\|
            \boldsymbol{s}_{\theta,a}(\tilde{\boldsymbol{R}})
            +
            \frac{\Delta \boldsymbol{r}_a}{\sigma_{\log P}^2}
        \right\|^2
    \right],
\end{equation}
which is the mean squared error between the predicted score
$\boldsymbol{s}_{\theta,a}(\tilde{\boldsymbol{R}})$ and the scaled denoising target
$-\Delta \boldsymbol{r}_a / \sigma_{\log P}^2$. Here, $\sigma_{\log P}=0.15$~\AA\ is a fixed scale parameter controlling the magnitude of the learned score field, independent of the sampled noise level $\sigma_n$. Thus, this is not a noise-conditional score model: the sampled $\sigma_n$ broadens the training distribution, while $\sigma_{\log P}$ fixes the scale of the deterministic denoising update.

Locally around a given prototype $c$, this Gaussian corruption model and score-matching objective encourage the network to approximate the conditional distribution of atomic displacements $\Delta\boldsymbol{r}_a = \boldsymbol{r}_a - \boldsymbol{R}_{0,a}^{(c)}$ as a Gaussian. In the simplest isotropic approximation,
\begin{equation}
\log \hat{P}_{\theta,c}(\boldsymbol{r}) \approx \mathrm{const}_c - \frac{1}{2\sigma_c^2} \sum_a \left\lVert \boldsymbol{r}_a - \boldsymbol{R}_{0,a}^{(c)} \right\rVert^2,
\end{equation}
so that the per-atom logit for the correct class $c^\star$ behaves as
\begin{equation}
l_{a c^\star} \approx \mathrm{const}_{c^\star} - \frac{1}{2\sigma_{c^\star}^2} \left\lVert \boldsymbol{r}_a - \boldsymbol{R}_{0,a}^{(c^\star)} \right\rVert^2.
\end{equation}
Thus, up to an additive constant and a phase-dependent scale, $l_{a c^\star}$ is proportional to the negative squared distance between the current atomic position and the ideal reference position for phase $c^\star$. This provides an approximate physical interpretation of the logit-based OPs as distance-like measures of similarity to each prototype. As shown for noisy Ag in the A\_hP2\_194 prototype in Fig.~\ref{fig:denoising_classification}b, the learned logits indeed follow an approximately linear relation with both the mean squared displacement and the input noise variance, consistent with this local Gaussian picture.

Model parameters are optimized using AdamW with a learning rate of $1\times10^{-3}$ and weight decay $1\times10^{-4}$. For small datasets (fewer than 50 structures), training the $\log \hat{P}_\theta$ decoder on top of a frozen MACE-MP backbone typically converges within about 2~hours on 4 nodes (4 GPUs per node). Across the full curated dataset, models initialized from MACE-MP weights show improved stability and slightly better performance in the score-matching objective, but do not consistently accelerate convergence of the full log-probability objective (Fig.~\ref{fig:pretrain_scratch_comparison_supp}), with the benefit most pronounced for larger MACE configurations (e.g., hidden irreps $128\times(0e+1o+2e)$ and 2 interaction layers). We attribute this to reusing the pretrained MACE-MP representation, including the scale and shift terms in the \texttt{ScaleShiftBlock}, which improves optimization stability and data efficiency. Overall, initialization from MACE-MP provides modest improvements in stability and denoising performance, but is not essential for training the log-probability model.

Training is performed with a batch size of 1 and a cosine learning-rate schedule. Training is continued until convergence, with the final model selected based on validation loss. The classification loss weight is set to $w_{\text{cl}} = 20$. The dataset is randomly split into 90\% training and 10\% validation sets using a fixed random seed.

\subsection*{Denoising and inference.}

At inference time, thermally perturbed configurations are treated as noisy inputs. Starting from an initial configuration $\boldsymbol{R}^{(0)}$, we iteratively apply the learned denoising field to generate
$\boldsymbol{R}^{(0)}, \boldsymbol{R}^{(1)}, \ldots, \boldsymbol{R}^{(T)}$.
Unless otherwise stated, we use $T=8$ denoising steps.

Atomic positions are updated deterministically using the predicted score field and the fixed scale parameter used during training,
\begin{align}
\boldsymbol{R}^{(t+1)}
=
\boldsymbol{R}^{(t)}
+
\sigma_{\log P}^{2}
\boldsymbol{s}_\theta(\boldsymbol{r}^{(t)}).
\end{align}

Phase labels are assigned using $\arg\max_c l_{ac}$, and the logits $l_{ac}$ are retained as continuous, phase-resolved OPs. We also report step-0 predictions, corresponding to classification without denoising, to isolate the effect of iterative structural refinement. Step-0 logits characterize the input configuration, whereas logits after denoising characterize the projection of that configuration onto the learned prototype manifold.

\subsection*{Confidence and ambiguity measures.}

To quantify the confidence of phase assignments, we define the logit margin for each atom $a$ as
\begin{equation}
m_a = l_{a,c_1} - l_{a,c_2},
\end{equation}
where $c_1$ and $c_2$ are the top-1 and top-2 predicted classes, respectively. A larger margin indicates a more confident assignment.

We also define the softmax entropy
\begin{equation}
S_a = -\sum_c p_{ac} \log p_{ac}, \quad
p_{ac} = \frac{\exp(l_{ac})}{\sum_{c'} \exp(l_{ac'})},
\end{equation}
which measures the uncertainty over phase assignments.

We note that the logits $l_{ac}$ are raw, uncalibrated scores; the margin and entropy therefore provide relative, model-internal measures of confidence rather than calibrated probabilities. Unless otherwise stated, ambiguity visualizations use the margin $m_a$, with lower margins indicating stronger competition between prototype assignments.

\section*{Results}

\subsection*{Model performance on large crystalline dataset}

We first assess the performance of the log-probability model on the curated Materials Project dataset\cite{jain2013commentary} described in the Methods. We evaluate the model on several datasets including Materials Project test structures\cite{jain2013commentary}, DC3 high-temperature configurations, and out-of-equilibrium systems such as shock-loaded Ti and water–ice interfaces. A summary of interpolation and extrapolation regimes is provided in Table~\ref{table:training_test_ood}. We also evaluate the model beyond the perturbation regimes used during training, including physically meaningful extrapolation tests involving noise amplitudes beyond the training range, high-temperature disorder and defects, liquid phases not seen during training, and large non-equilibrium strains. The model is trained jointly for denoising and crystalline prototype classification: given a noisy atomic configuration, it predicts per-atom, per-phase logits (unnormalized scores prior to normalization) and a conservative score field whose gradient is used to iteratively refine atomic positions (Figure~\ref{fig:workflow}). This shared log-probability landscape underlies both the denoising dynamics and the final phase assignments.

Table~\ref{tab:accuracy} verifies that the learned log-probability landscape preserves prototype identity on curated crystalline structures and synthetic perturbations, including ice polymorphs, elemental, binary, and ternary compounds spanning hundreds of AFLOW prototypes. The high accuracies indicate consistency with the prototype labels used for training and validation, while denoising errors remain sub-\AA. For the combined elemental+binary set (7{,}746 structures, 403 structure types), the model reaches classification accuracies above 99.9\% on clean inputs and maintains similarly high accuracy on Gaussian-perturbed structures with a noise standard deviation of 0.15~\AA, while keeping the denoising RMSE below 0.002~\AA. Notably, the chemistry-agnostic elemental model---which shares a single ML representation across all elements---still attains $\sim$96\% accuracy, indicating that the ML descriptors capture robust geometric information even without explicit chemical labels.

The chemistry-agnostic model may also be useful for extending the framework to compositionally complex systems such as high-entropy alloys, where many elements can share similar lattice environments.

\begin{table}[h!]
\centering
\begin{tabular}{|>{\centering\arraybackslash}p{2cm}|>{\centering\arraybackslash}p{1cm}|>{\centering\arraybackslash}p{1cm}|>{\centering\arraybackslash}p{1cm}|>{\centering\arraybackslash}p{3cm}|>{\centering\arraybackslash}p{3cm}|>{\centering\arraybackslash}p{3cm}|}
\hline
\textbf{Material}&\textbf{\#  structures}&\textbf{\# prototypes}&\textbf{\# atom types}&\textbf{Class. acc. at step~8 (clean / perturbed 0.15~\AA)}&\textbf{RMSE (\AA)}&\textbf{Class. acc. at step~0} \\ \hline
Ice  & 7 & 7 & 2 &1.0000 / 1.0000 & 0.0191 / 0.0191 &  1.0000 / 1.0000     \\ \hline
Elemental structures & 238 & 33 & 72 & 0.9961 / 1.0000  & 0.0002 / 0.0013 & 0.9961 / 0.9961\\ \hline
Binary structures & 7488 & 363 & 75 & 0.9988 / 0.9983 & 0.0013 / 0.0019 & 0.9991 / 0.9991 \\ \hline
Ternary structures & 14848 & 373 & 84 & 0.9977 / 0.9937 & 0.0019 / 0.0020 & 0.9981 / 0.9972 \\ \hline
Elemental + binary structures & 7746 & 403 & 75 & 0.9993 / 0.9991 & 0.0009 / 0.0011 & 0.9994 / 0.9990 \\ \hline
Elemental chemistry-agnostic & 238 & 33 & 1 & 0.9625 / 0.9595 & 0.0054 / 0.0091 & 0.9628 / 0.9699 \\ \hline
\end{tabular}
\caption{Performance of the log-probability model on the curated Materials Project dataset. The model jointly learns to denoise atomic coordinates and classify crystal prototypes across ice polymorphs, elemental, binary, and ternary compounds. For each dataset, we report the number of structures, prototypes, and atom types, together with classification accuracy on clean and perturbed inputs (Gaussian noise up to 0.15~\AA), denoising RMSE, and accuracy at step~0 (before any denoising steps are applied). The high accuracies verify consistency with the curated prototype labels, while the low RMSE values verify the learned denoising field on synthetic perturbations.}
\label{tab:accuracy}
\end{table}

We leveraged the strong expressive power of the MACE-MP model by reusing its featurization layers and adding a new trainable decoder that predicts per-atom logits and the aggregated $\log \hat{P}_\theta$. This transfer-learning setup improves optimization stability and yields slightly better performance in the score-matching objective, while achieving comparable classification accuracy to training from scratch (see Fig.~\ref{fig:pretrain_scratch_comparison_supp} and Methods).

\subsection*{Multi-phase denoising and classification in ice polymorphs}

\begin{figure}
    \centering
    \includegraphics[width=1\linewidth]{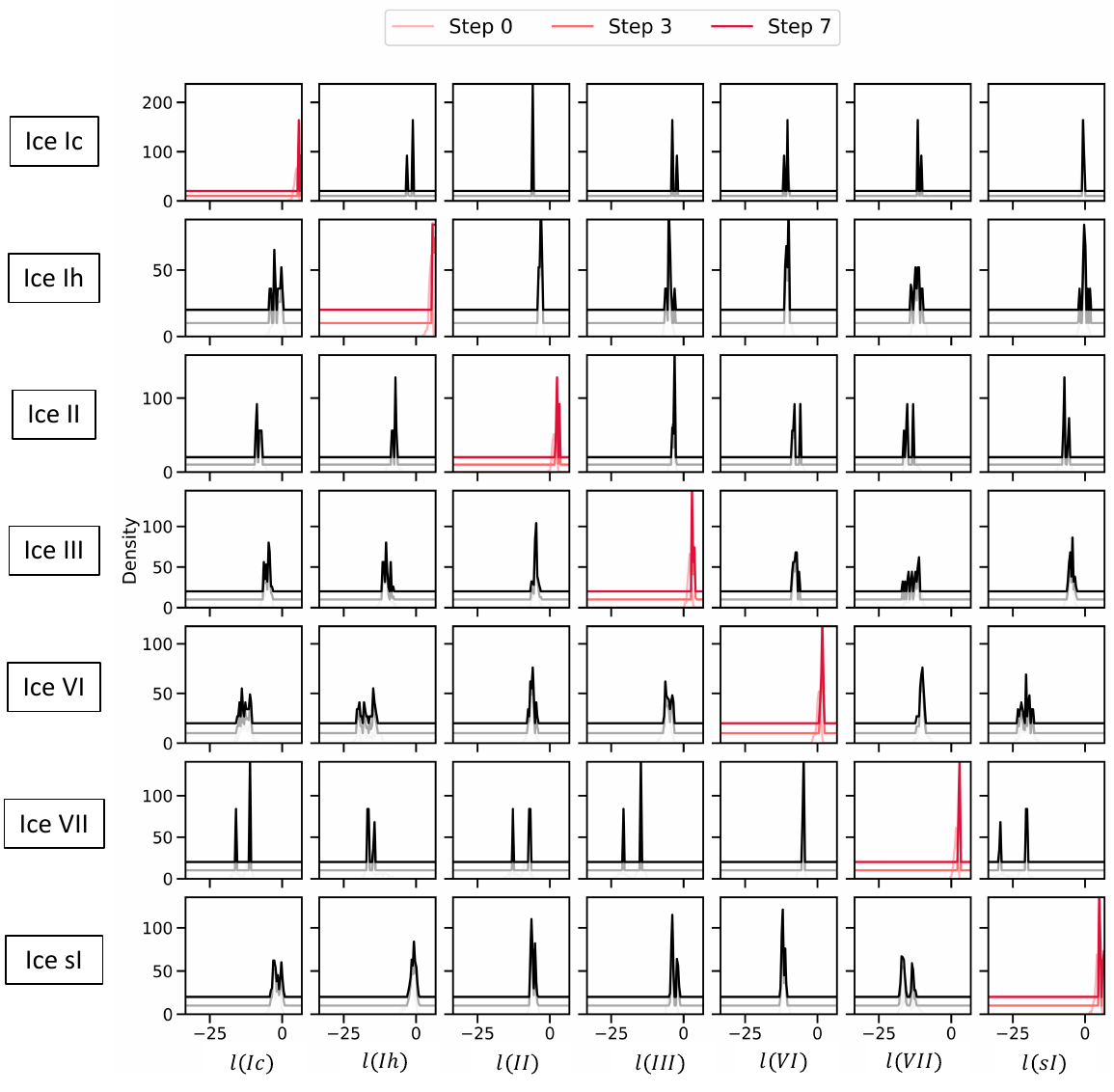}
    \caption{Per-phase logit distributions $l_{ac}$ for 7 ice polymorphs (I$_\mathrm{c}$, I$_\mathrm{h}$, II, III, VI, VII, and sI). Each row corresponds to a given true ice phase and each column to a predicted structural class. Within each panel, the light curves show the initial perturbed structures, intermediate curves show partially denoised configurations (after 3 of 8 denoising steps), and the darkest curves show the fully denoised structures. Diagonal panels (true class = predicted class) develop sharp, high-$l$ peaks as denoising proceeds, indicating confident and self-consistent phase recognition, while off-diagonal panels remain suppressed at low $l$.}
    \label{fig:ice}
\end{figure}

As a first multi-phase test, we apply the model to 7 ordered ice polymorphs (I$_\mathrm{c}$, I$_\mathrm{h}$, II, III, VI, VII, and sI), which provide a familiar but nontrivial benchmark with distinct hydrogen-bonding networks and local environments. The model is trained jointly on all 7 phases and evaluated on Gaussian-perturbed structures with noise amplitudes up to $\sigma_{\max}=0.15$~\AA, using the same denoising protocol as for the crystalline solids. The Gaussian displacements mimic thermal-like positional fluctuations around the ideal lattice sites.

Figure~\ref{fig:ice} shows the distributions of per-atom logits $l_{ac}$ for each input phase (rows) and predicted structural class (columns), at different stages of the denoising process. Light-colored curves correspond to the initial perturbed configurations, intermediate curves correspond to partially denoised structures (e.g., after 3 out of 8 denoising steps), and the darkest curves represent the fully denoised outputs. Along the diagonal panels--where the predicted class matches the true phase--the $l_{ac}$ distributions develop pronounced peaks at high values as denoising proceeds, indicating confident and self-consistent classification. Off-diagonal panels remain narrowly peaked at lower $l_{ac}$, reflecting smaller weights assigned to incorrect phases.

Quantitatively, the model achieves perfect classification accuracy (1.000) for all seven ice phases, both for clean inputs and for perturbed structures with $\sigma_{\max}=0.15$~\AA, while maintaining denoising RMSEs on the order of $2\times 10^{-2}$~\AA\ (Table~\ref{tab:accuracy}). These results demonstrate that a single probabilistic model can robustly distinguish multiple hydrogen-bonded phases even under substantial thermal-like perturbations. They also illustrate how the per-phase logits $l_{ac}$ naturally act as continuous OPs: each phase is associated with a distinct, well-separated logit distribution that sharpens under denoising, providing a scalar measure of structural similarity suitable for tracking phase identity and transformation pathways. In contrast, a separate, second-stage descriptor-based classifier was needed to supplement the non-conservative denoiser in Ref.~\cite{Sun2024JCIM-ice}.

\subsection*{Interpretable OPs and continuous transformation paths}

We next examine how the model behaves on familiar close-packed structures and along continuous deformation paths between them. This serves both as a sanity check that the machine-learned $\log \hat{P}_\theta$ landscape respects well-known crystallographic relationships and as a quantitative test of the physical interpretability of the logit-based OPs.

\begin{figure}
    \centering
    \includegraphics[width=\linewidth]{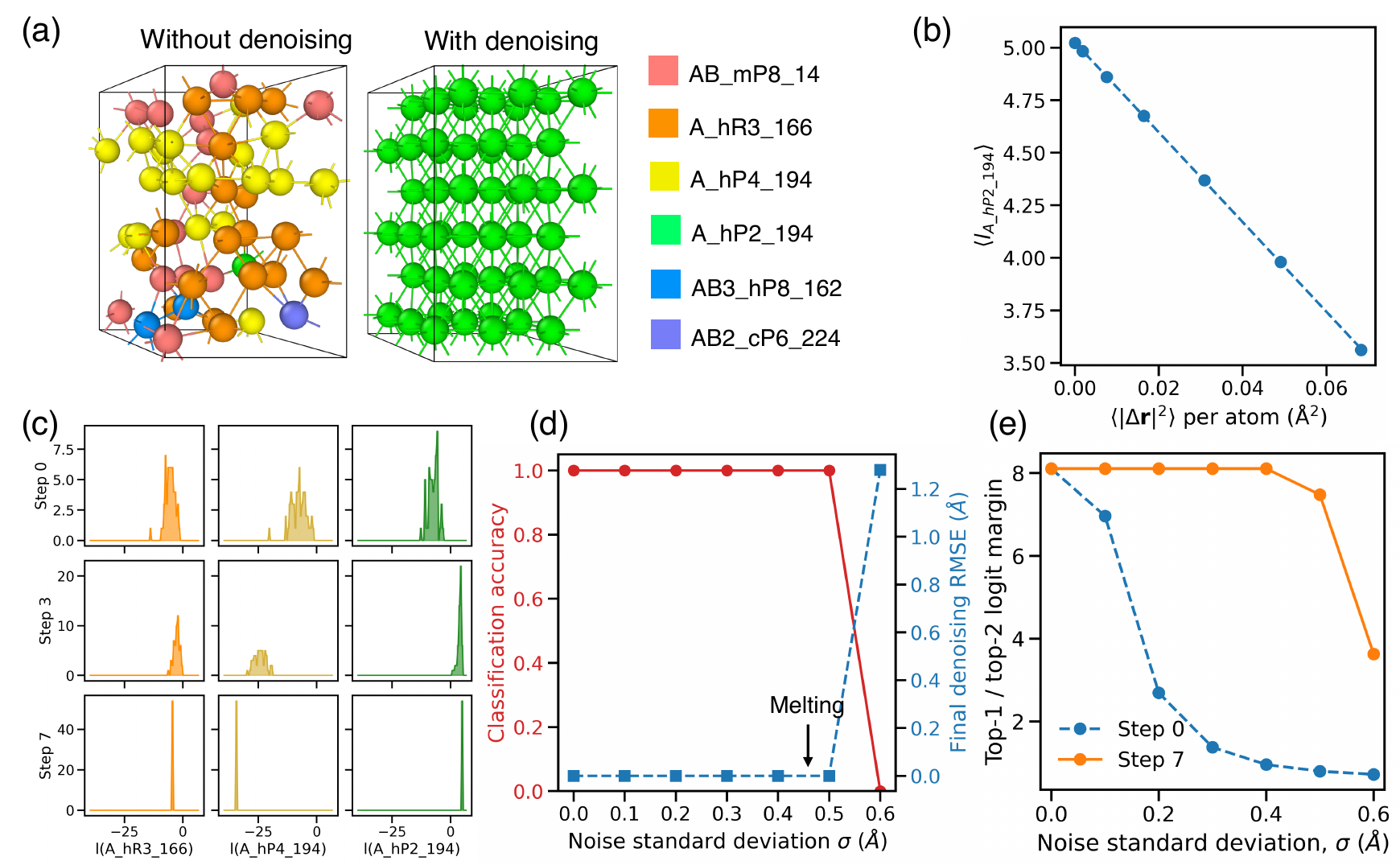}
    \caption{Interplay between denoising, classification, and logit-based OPs for noisy Ag in the A\_hP2\_194 prototype. (a) Example with strong Gaussian noise ($\sigma = 0.4$~\AA). In classification-only mode (no denoising), the structure is misclassified into several competing AFLOW prototypes. When denoising and classification are coupled through the log-probability model, the atomic positions are iteratively refined toward high-$\log \hat{P}_\theta$ regions and the correct A\_hP2\_194 label is recovered for all atoms. (b) Mean logit for the correct prototype, $\langle l(\mathrm{A\_hP2\_194})\rangle$, versus mean squared displacement per atom, $\langle|\Delta \boldsymbol{r}|^2\rangle$, for a range of initial noise levels and denoising steps. The approximately linear trends (dashed regression lines) are consistent with the local Gaussian model in which $l$ is proportional to the negative squared distance to the ideal structure, providing an approximate physical interpretation of the logit-based OP. (c) Evolution of per-phase logit distributions for 3 closely related prototypes, $l(\mathrm{A\_hP3\_166})$, $l(\mathrm{A\_hP4\_194})$, and $l(\mathrm{A\_hP2\_194})$, at denoising steps 0, 3, and 7. As denoising proceeds, the logit distribution for the correct A\_hP2\_194 phase sharpens and shifts to higher values, while competing phases are suppressed, illustrating how the logit-based OPs become more decisive as the structure is projected onto the high-probability manifold. (d) Classification accuracy and final denoising RMSE as a function of the initial Gaussian noise standard deviation. The model maintains 100\% accuracy up to $\sigma \approx 0.5$~\AA, beyond which both accuracy and denoising quality degrade as the structures melt and no longer correspond to well-defined crystalline phases. (e) Noise-dependent logit margin (difference between the top-1 and top-2 logits) before (step 0) and after denoising (step 7). The margin decreases monotonically with increasing noise prior to denoising, indicating growing ambiguity between competing structural prototypes. After denoising, large margins are largely restored up to $\sigma \approx 0.5$~\AA, while at $\sigma=0.6$~\AA\ the margin remains reduced, signaling an out-of-manifold regime.}

    \label{fig:denoising_classification}
\end{figure}

Figure~\ref{fig:denoising_classification} focuses on an Ag structure in the hexagonal A\_hP2\_194 (space group 194) prototype and is intentionally designed to probe extrapolation beyond the training perturbation regime of the coupled denoising--classification inference. While the model is trained with Gaussian noise whose standard deviation is drawn uniformly up to $\sigma_{\max}=0.15$~\AA\ (Methods), here we evaluate substantially larger perturbations, including $\sigma=0.4$~\AA, to assess whether the learned $\log \hat{P}_\theta(\boldsymbol{r})$ landscape still provides a meaningful restoring drive toward the prototype manifold.

Panel~\ref{fig:denoising_classification}a illustrates the qualitative difference between classification-only inference and the coupled denoising+classification inference for this strongly perturbed input ($\sigma=0.4$~\AA). In classification-only mode (no denoising), the distorted local environments yield weak separation among competing prototypes, so no single class is strongly favored; consequently, atoms are distributed across multiple AFLOW labels. This behavior reflects a low-confidence near-tie regime rather than a confident but incorrect decision: at step~0 the per-phase logits occupy similar ranges and exhibit substantial overlap (panel~\ref{fig:denoising_classification}c, top row). In contrast, when denoising is enabled, atomic positions are iteratively updated using the conservative score field $\boldsymbol{s}_\theta(\boldsymbol{r})=\nabla_{\boldsymbol{r}}\log \hat{P}_\theta(\boldsymbol{r})$ (Eq.~\ref{eq:score}), which drives the configuration toward higher-$\log \hat{P}_\theta$ regions and yields a self-consistent recovery of the A\_hP2\_194 assignment.

Panel~\ref{fig:denoising_classification}b directly probes the approximate quadratic relation between the logits and the displacement from the ideal reference structure derived in the Methods section. It plots the mean logit for the correct phase, $\langle l(\mathrm{A\_hP2\_194})\rangle$, versus the mean squared displacement per atom, $\langle|\Delta \boldsymbol{r}|^2\rangle$, along the denoising trajectory for multiple initial noise levels. The data align closely with a linear trend (shown by a regression line), consistent with the score-matching formulation (Method).
\begin{align}
l \approx \mathrm{const} - \lVert \Delta \boldsymbol{r}  \rVert^2/(2\sigma^2)
=
\mathrm{const} - \lVert \boldsymbol{r} - \boldsymbol{R}_0\rVert^2/(2\sigma^2)
\end{align}
relative to the correct ideal phase $\boldsymbol{R}_0$. The plot provides explicit evidence that the learned logits admit an approximate physical interpretation as distance-like OPs measuring proximity to the corresponding ideal prototype.

Panel~\ref{fig:denoising_classification}c shows how the per-phase logit distributions evolve during denoising for 3 closely related close-packed prototypes, $l(\mathrm{A\_hP3\_166})$, $l(\mathrm{A\_hP4\_194})$, and $l(\mathrm{A\_hP2\_194})$, at denoising steps 0, 3, and 7. At step~0, the noisy structure exhibits broad, partially overlapping logit distributions, and the correct A\_hP2\_194 class is not clearly dominant. After a few denoising steps, the logit distribution for A\_hP2\_194 sharpens and shifts to higher values, while the competing phases are suppressed and pushed toward lower logits. By step~7, the correct class forms a well-separated high-$l$ peak, and the impostor phases remain narrowly distributed at low $l$. This illustrates how the logit-based OPs act as continuous, phase-resolved measures of structural similarity that naturally become more decisive as the structure is projected onto the learned high-probability manifold.

Panel~\ref{fig:denoising_classification}d summarizes the net effect on predictive performance by plotting the classification accuracy and final denoising RMSE as functions of the initial Gaussian noise standard deviation. The model maintains 100\% classification accuracy for perturbations up to 0.5~\AA, with small denoising errors, and both metrics degrade beyond this point as the structures melt and no longer correspond to well-defined crystalline phases. Together, panels a--d show that the $\log \hat{P}_\theta$ model not only stabilizes classification through denoising but also yields logit-based OPs that vary smoothly and approximately quadratically with the squared distance to the underlying prototype, in line with the intended probabilistic interpretation.

To further quantify how ambiguity emerges under strong perturbations, panel~\ref{fig:denoising_classification}e shows the noise dependence of the logit margin, defined as the difference between the top-1 and top-2 logits. Prior to denoising (step 0), the margin decreases monotonically with increasing noise, reflecting growing competition between structural prototypes. After denoising, large margins are restored up to $\sigma \approx 0.5$~\AA, indicating confident identification of the correct structure. At $\sigma=0.6$~\AA, however, the margin remains reduced, indicating a loss of structural identifiability. This behavior supports the interpretation of the logits as continuous measures of structural similarity, with the margin acting as a soft OP that quantifies the degree of structural ambiguity or departure from the prototype manifold.

\begin{figure}
    \centering
    \includegraphics[width=\linewidth]{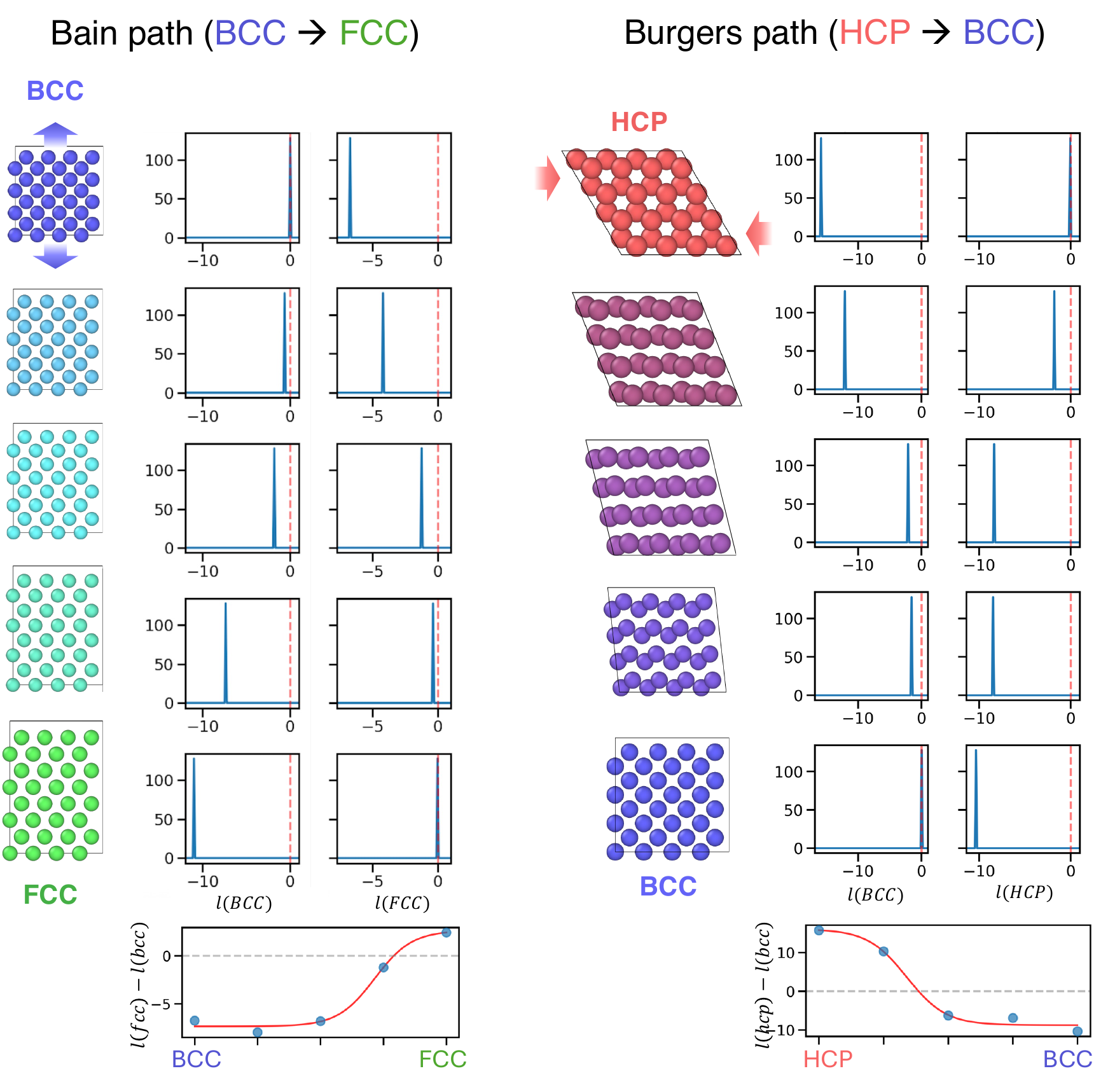}
    \caption{Evolution of logit-based OPs along continuous transformation paths. The model is evaluated along (left) the Bain path connecting BCC and FCC and (right) the Burgers path connecting HCP and BCC, using sequences of intermediate structures. For each configuration, per-phase logits for the competing structures are evaluated and aggregated into prototype-resolved OPs. Along the Bain path, $l(\mathrm{BCC})$ decreases while $l(\mathrm{FCC})$ increases, crossing smoothly near the midpoint; along the Burgers path, $l(\mathrm{HCP})$ decays as $l(\mathrm{BCC})$ rises. The smooth exchange of logit-based OP weight between phases shows that the model captures continuous structural evolution rather than treating prototypes as isolated categories.}
    \label{fig:continuous}
\end{figure}

To probe whether the $\log \hat{P}_\theta$ model captures smooth structural evolution between phases, we evaluate it along two standard transformation paths: the Bain path connecting BCC and FCC, and the Burgers path connecting HCP and BCC (Figure~\ref{fig:continuous}). Along each path, we generate a sequence of intermediate configurations with gradually changing lattice parameters and atomic positions. For each configuration, we evaluate the per-atom, per-phase logits $l_{ac}$ and aggregate them into prototype-resolved OPs.

The resulting profiles of these per-phase logits and their differences demonstrate their usefulness as continuous, physically interpretable OPs. Along the Bain path, the BCC logit-based OP starts high in the initial BCC-like region and decreases monotonically as the structure is distorted toward FCC, while the FCC logit-based OP rises in a complementary fashion and dominates near the FCC endpoint. Similarly, along the Burgers path, the HCP logit-based OP decreases as the structure is driven toward BCC, whose logit-based OP increases and eventually becomes dominant. This smooth exchange of OP weight between competing phases indicates that the model does not treat prototypes as discrete, disconnected categories, but instead learns a continuous OP landscape over configuration space that tracks gradual structural transformations. Having emerged naturally from the crystalline structures alone, without requiring access to the underlying physics (e.g.\ an energy landscape) or detailed chemistry, these OPs can be defined in a consistent and universal way with a direct physical meaning related to the squared distance to the corresponding ideal phases.

\subsection*{Robustness to thermal disorder and point defects}

We next evaluate robustness under realistic thermal disorder and local defects, where
traditional template- and threshold-based structure identifiers often struggle. For thermal effects, we use the DC3 database~\cite{chung2022data}, which contains molecular dynamics (MD) snapshots of elemental and binary crystals equilibrated at high temperatures near  their melting points. These configurations exhibit large vibrational amplitudes and, importantly, can also contain non-thermal disorder such as vacancies, interstitials, and stacking faults. Such environments are challenging for hard local classifiers because the local neighbor topology is no longer well represented by an ideal lattice template. Compared with our previous denoising-only workflow, in which a score-based denoiser was followed by external classifiers such as PTM or CNA, the present model performs denoising and prototype assignment within the same log-probability framework. We therefore evaluate the highest-temperature available DC3 snapshots as a stringent test of direct logit-based classification under strong thermal disorder~\cite{Hsu2024nCM-denoiser}.

Figure~\ref{fig:DC3} compares classification performance of the log-probability model against two widely used baselines, PTM and CNA. For each DC3 system, we take the highest-temperature snapshot available and apply $k=0,\dots,8$ denoising steps using the model (with $k=0$ corresponding to the original DC3 snapshot). Thus, unlike the previous denoising-plus-classifier pipeline, the present model is applied without system-specific retraining or a separate downstream classifier, and its step-0 predictions directly test the transferability of the learned logit representation under strong thermal disorder. At each step $k$, we evaluate all three methods on the same coordinates, i.e.\ on the configuration obtained after $k$ denoising steps. Across most tested systems, the model attains higher accuracy with fewer denoising iterations than PTM or CNA, and in many cases reaches perfect phase identification within a few steps even when the structures remain visibly noisy. This reflects a key difference in philosophy: PTM and CNA rely on discrete, hand-crafted neighbor and topology criteria tuned to ideal lattices, whereas the model learns a probabilistic association between a broad distribution of thermally perturbed local environments and their corresponding prototypes.

\begin{figure}
    \centering
    \includegraphics[width=\linewidth]{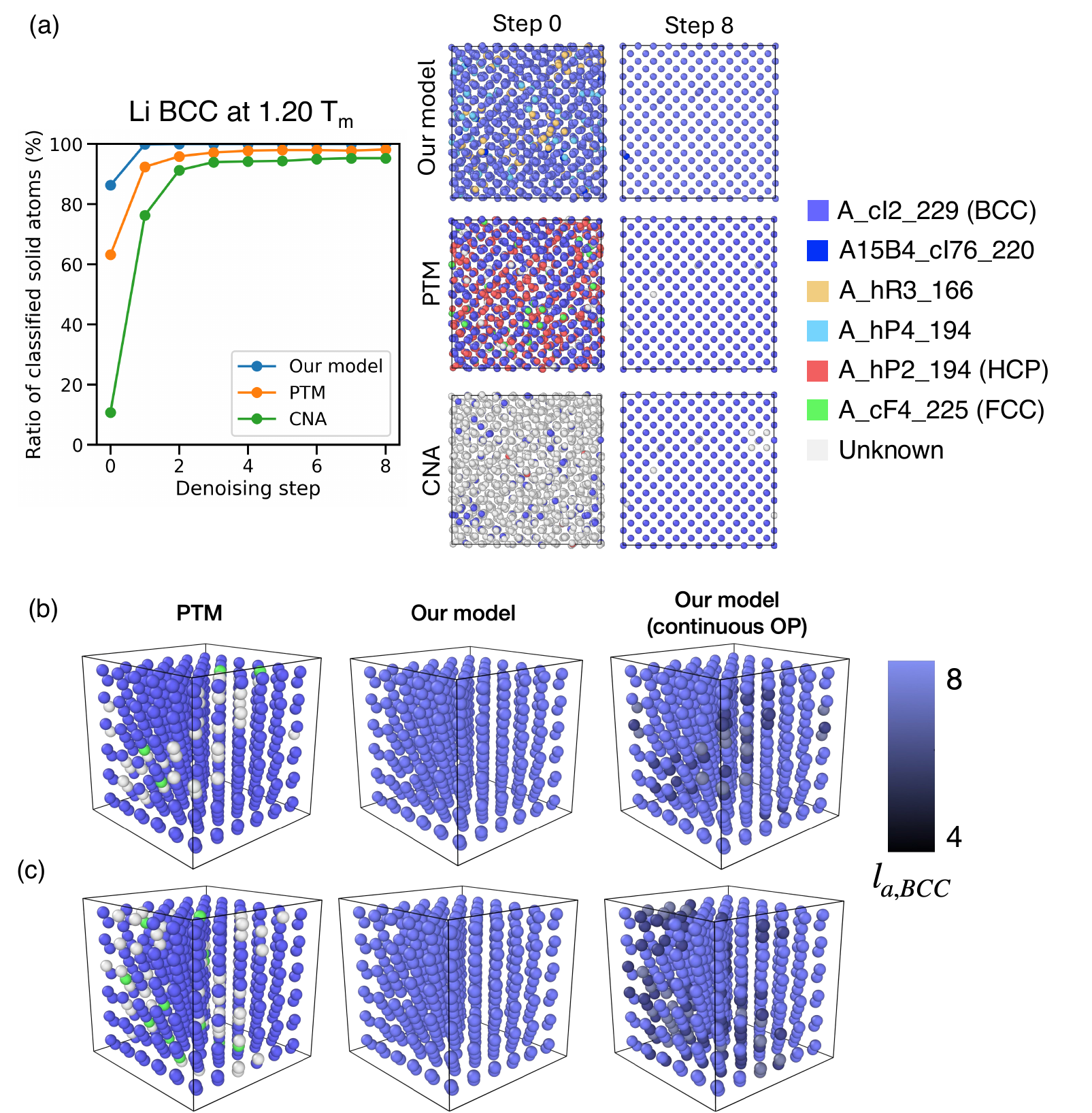}
    \caption{Robustness to thermal disorder and point defects. (a) Classification of BCC Li at $1.20~T_m$. The log-probability model achieves higher accuracy than PTM/CNA on the raw high-temperature snapshot. Applying log-probability denoising improves all methods when evaluated on the denoised coordinates, but PTM/CNA typically plateau below 100\% because vacancy/interstitial defects and other non-thermal disorder are preserved. (b,c) Defective BCC Fe with 5 and 10 vacancies (out of 432 atoms). PTM misclassifies atoms near vacancy cores as FCC or ``unknown,'' while the model assigns BCC via $\arg\max_c l_{ac}$. Coloring by $l_{a,\mathrm{BCC}}$ reveals defect neighborhoods as low-logit halos.}
    \label{fig:point_defect}
\end{figure}

As a representative example,  Fig.~\ref{fig:point_defect}a shows BCC Li at $1.20\,T_m$. On the raw snapshot ($k=0$), the model identifies BCC more  reliably than PTM/CNA. As denoising  proceeds, all methods improve when  evaluated on the same denoised coordinates, but PTM/CNA retain a small fraction of non-BCC labels even at late steps.

A natural question is why PTM/CNA do not always reach 100\% agreement with the reference label even after $k=8$ denoising steps for some systems (e.g., BCC Li/Fe). In addition to BCC Li, there are a few other cases in Fig.~\ref{fig:DC3} with inconsistent classifications even at step 8. The reason is the presence of defects, e.g.\ vacancies, interstitials and Frenkel pairs, in these high temperature structures above $T_m$. Log-probability denoising is designed to suppress the approximately Gaussian thermal component while preserving such physically meaningful defect cores; consequently, the local environments near defects can remain far from any ideal template and may be labeled as ``Other/Unknown'' (or occasionally as a nearby lattice type) depending on the thresholds of PTM/CNA (Fig.~\ref{fig:DC3}a).

Our model is not always the most accurate at very early denoising steps. In particular, for close-packed systems the few-step HCP accuracy can trail PTM. This is a consequence of the broader hypothesis space of the model: it predicts logits over many closely related close-packed AFLOW prototypes (differing by stacking variants and subtle long-range order), which can be nearly degenerate under strong thermal disorder at $k=0$ or $k=1$. PTM, by contrast, typically distinguishes only a small set of close-packed templates (most commonly FCC vs.\ HCP). In applications where only a few phases are physically relevant, this gap can be mitigated by running additional denoising steps or by restricting inference to a reduced  candidate prototype set.

To directly probe defect sensitivity, we introduce vacancy-type defects into a BCC Fe supercell by randomly removing a small fraction of atoms (5 and 10 vacancies out of 432 atoms, respectively; Figure~\ref{fig:point_defect}b and c). These missing atoms distort the local environments around the defect cores and frequently cause PTM to misclassify neighboring atoms as FCC or label them as ``unknown'' (left panels), reflecting the fragility of hard, template-based labels under local coordination changes. In contrast, the log-probability model correctly assigns all atoms to the BCC prototype for both vacancy concentrations (middle panels), preserving the global phase identity. At the same time, the continuous BCC logit-based OP provides a natural defect-sensitive measure of local order: when atoms are colored by their BCC logit value $l_{a,\mathrm{BCC}}$ (right panels), the undisturbed crystal interior appears uniformly bright (high $l_{a,\mathrm{BCC}}$), while shells surrounding the vacancies show localized depressions in $l_{a,\mathrm{BCC}}$ (darker purple), indicating reduced confidence and stronger local disorder. Thus, the model simultaneously maintains robust global phase recognition and yields a smooth, quantitative measure of local deviations from ideal BCC order that discrete template matching cannot provide.

\subsection*{Application to diverse binary prototypes}

While many structure-identification methods are tuned to a small set of familiar lattices (e.g., BCC, FCC, HCP), the AFLOW prototype library contains a much broader spectrum of low-symmetry and less common structures. To assess whether the model extends beyond close-packed metals and simple oxides, we evaluate it on binary systems with multiple polymorphs and nontrivial AFLOW labels.

\begin{figure}
    \centering
    \includegraphics[width=\linewidth]{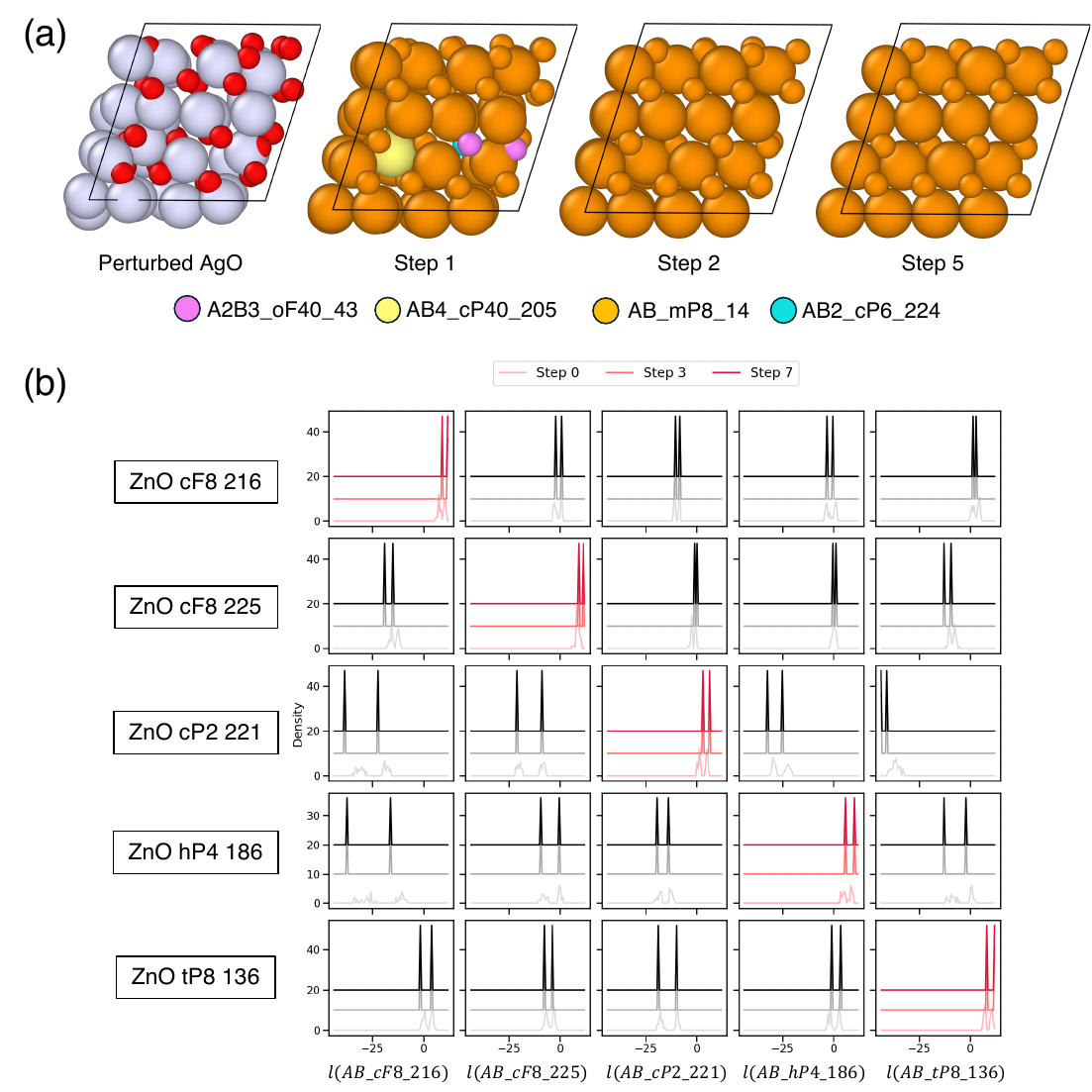}
    \caption{Application to diverse binary prototypes. (a) AgO in the AB\_mP8\_14 prototype: starting from a perturbed configuration, the model progressively denoises the structure while tracking per-atom prototype labels. At early steps, some atoms are transiently assigned to competing prototype classes (e.g., A2B3\_oF40\_43, AB4\_cP40\_205, AB2\_cP6\_224), but these inconsistencies vanish as denoising proceeds and all atoms converge to the correct AB\_mP8\_14 class. (b) 5 ZnO polymorphs: for each prototype (rows), the evolution of per-atom logit-based OP distributions across denoising steps (columns) shows sharpening, well-separated peaks for the true class and suppressed values for competing classes. These examples highlight that the approach is not limited to simple BCC/FCC/HCP lattices but extends to low-symmetry AFLOW prototypes in binary systems.}
    \label{fig:binary structure}
\end{figure}

Figure~\ref{fig:binary structure} illustrates two representative examples. Panel~\ref{fig:binary structure}a shows an AgO structure in the AB\_mP8\_14 prototype, starting from a perturbed configuration and followed through successive denoising steps. At early iterations (e.g., step~1), some atoms are transiently assigned to alternative prototype classes such as A2B3\_oF40\_43, AB4\_cP40\_205, or AB2\_cP6\_224, reflecting local environments that momentarily resemble competing motifs. As denoising proceeds, these inconsistencies vanish and the model converges to a self-consistent assignment in which all atoms are correctly classified as the target AB\_mP8\_14 prototype.

Panel~\ref{fig:binary structure}b considers 5 distinct ZnO polymorphs. For each prototype (rows), we track the evolution of per-atom logit distributions across denoising steps (columns). The light curves correspond to the initial perturbed structures, intermediate curves show partially denoised states (step~3 of 8), and the darkest curves represent the fully denoised outputs. In all cases, the logit distributions for the true prototype sharpen into a dominant, well-separated peak, while the competing classes remain suppressed. Together, these examples demonstrate that the log-probability model is not restricted to a small set of canonical lattices, but readily generalizes to diverse, low-symmetry AFLOW prototypes in binary systems.

\subsection*{LogP OPs in mixed solid--liquid water-ice systems}

\begin{figure}
    \centering
    \includegraphics[width=\linewidth]{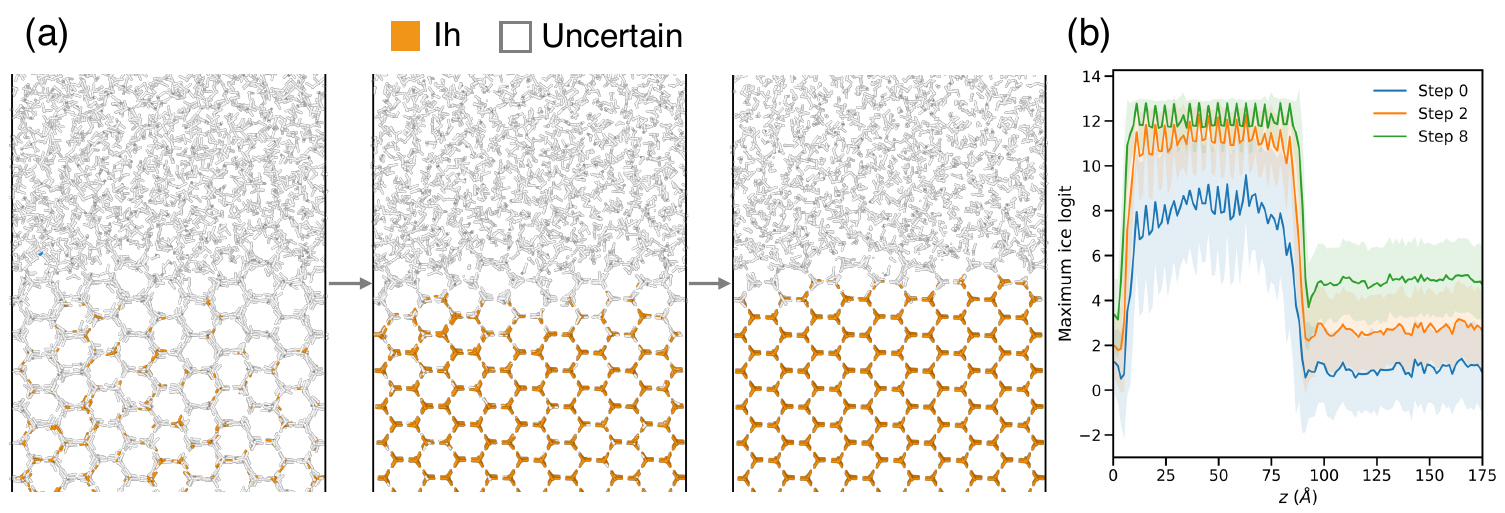}
    \caption{Probabilistic order parameters at a water--ice interface. (a) Denoising trajectory and probabilistic classification for a finite-temperature water--ice coexistence configuration. The left panel shows the initial thermally perturbed configuration, where a subset of interfacial and crystalline environments remain structurally ambiguous (white) due to small top-2 logit margins. The middle and right panels correspond to intermediate and late denoising steps, respectively. As denoising progresses, the crystalline ice region relaxes toward a more coherent high-log-probability manifold, progressively sharpening the distinction between crystalline and liquid-like environments and substantially reducing ambiguous classifications near the interface. Orange denotes Ih-like environments and white denotes atoms with small top-2 logit margins. (b) Spatial profile of the maximum ice logit OP, $\max_i l_{a,i}$ where the maximum is taken over all trained ice polymorph logits. The crystalline region exhibits uniformly large values corresponding to strongly ice-like environments, whereas the liquid region remains substantially lower and spatially diffuse, yielding a smooth probabilistic transition across the solid--liquid interface. Shaded regions indicate the standard deviation within each spatial bin.}
    \label{fig:water example}
\end{figure}

To probe the behavior of the log-probability model in heterogeneous environments containing both ordered and disordered regions, we apply it to a water--ice coexistence system obtained from molecular dynamics simulations at 300~K and 1~kbar (Figure~\ref{fig:water example}). The model is trained on 7 ordered ice polymorphs (Ic, Ih, II, III, VI, VII, and sI), while liquid water is not included during training and therefore represents a structurally distinct disordered environment expected to remain low in all ice-related logits. Thermal fluctuations and interfacial disorder naturally introduce substantial positional noise throughout the system, making the distinction between crystalline and liquid-like environments particularly challenging near the interface. This setting provides a stringent test for local order parameters because finite-temperature fluctuations continuously distort the local hydrogen-bond network and partially blur the crystalline manifold.

Figure~\ref{fig:water example}a shows the evolution of probabilistic classification during iterative denoising. In the initial thermally perturbed configuration, the distinction between crystalline ice and liquid water is not sharply resolved: many atoms have relatively low maximum ice logits or small top-2 logit margins, so a substantial fraction of environments are classified as ambiguous. After two denoising steps, the crystalline region begins to emerge more clearly, especially away from the interface, but interfacial environments remain diffuse. By the final denoising step, the ice slab forms a coherent Ih-like region with a well-defined boundary, while the liquid region remains low-confidence and structurally disordered. Thus, denoising progressively sharpens the crystalline manifold and converts a thermally blurred interface into a clearer probabilistic solid--liquid transition.

This behavior is quantified in Figure~\ref{fig:water example}b using a probabilistic ice-likeness OP defined as the maximum logit over all trained ice polymorphs, $\max_i l_{a,i}$. At step 0, the OP remains relatively low and broadly distributed throughout the system due to strong thermal disorder in the finite-temperature configuration. As denoising proceeds, the OP increases substantially within the crystalline region and develops a clear high-value plateau corresponding to strongly ice-like environments, while the liquid region remains substantially lower and spatially heterogeneous. Importantly, the liquid region does not collapse into a crystalline phase during denoising, but instead remains low in all ice-related logits, indicating that the model preserves structurally disordered environments outside the crystalline manifold. Also, the transition across the interface emerges continuously rather than through a discrete structural threshold, allowing partially ordered and interfacial environments to be represented smoothly within the same probabilistic framework.

Unlike geometric template-matching approaches that rely on hard structural criteria, the log-probability framework naturally provides both probabilistic phase assignment and a relative ambiguity measure through the logit margin. Additional comparisons against conventional structural analysis methods, including OVITO CHILL+, are provided in the Supporting Information (Fig.~\ref{fig:water_chill_compare}). These results demonstrate that log-probability-based order parameters can robustly characterize thermally fluctuating mixed-phase systems while simultaneously enabling denoising, classification, and continuous interfacial order characterization within a unified framework. This example highlights how per-atom logit-based OPs derived from the log-probability model can serve as probabilistic OPs for complex interfacial systems, with potential applications to solid--liquid coexistence, nucleation, and interfacial free-energy estimation~\cite{hoyt2001method}.

\subsection*{Shock-compressed Ti: extrapolation to highly nonequilibrium structures}

\begin{figure}
    \centering
    \includegraphics[width=\linewidth]{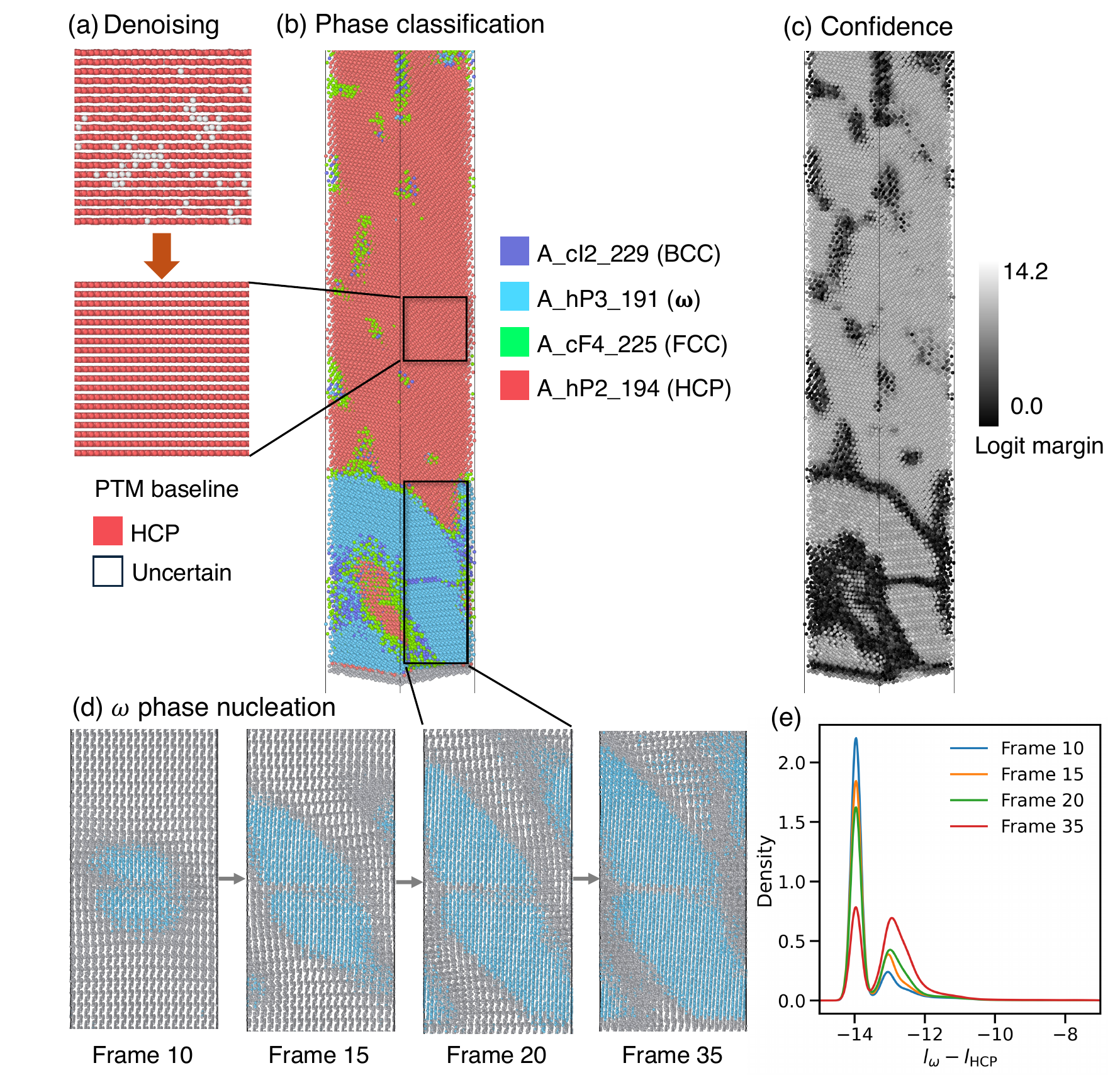}
    \caption{Shock-compressed Ti as a comprehensive extrapolation test. (a) Example of local denoising on a shock-compressed HCP Ti region. PTM applied directly to the noisy configuration leaves many atoms unassigned (white), whereas the denoised configuration restores coherent local ordering. (b) Spatial phase classification obtained from the log-probability model for frame 75, showing coexisting HCP-like and $\omega$-like regions under strong deformation. (c) Spatial map of the top-2 logit margin, $l_{(1)}-l_{(2)}$, where $l_{(1)}$ and $l_{(2)}$ are the largest and second-largest logits, respectively. Large margins indicate confident structural assignments, while small margins identify structurally ambiguous regions such as interfaces and highly distorted environments. (d) Evolution of $\omega$ domains during shock propagation (frames 10, 15, 20, and 35), illustrating nucleation and growth of the $\omega$ phase. Only $\omega$-classified atoms are highlighted. (e) Distribution of the differentiable structural order parameter $l_{\omega}-l_{\mathrm{HCP}}$ for the corresponding frames. The progressive shift of the distribution toward larger values indicates the increasing population of $\omega$-like local environments during the HCP$\rightarrow\omega$ transformation.
}
    \label{fig:babak structure-2}
\end{figure}

As a final extrapolation test to show how the $\log P$ model can be applied in a real-world multi-task setting, we apply the log-probability model to a large-scale simulation of shock-compressed Ti containing severe deformation, phase coexistence, and highly nonequilibrium local environments (Figure~\ref{fig:babak structure-2}). The shocked structure contains coexisting HCP-like and $\omega$-like domains together with highly distorted interfacial regions, providing a challenging benchmark for generalized structural recognition under dynamic loading conditions.

Figure~\ref{fig:babak structure-2}a demonstrates the effect of the denoising step on a representative shock-compressed HCP Ti region. Direct application of PTM (RMSD cutoff = 0.07~\AA) to the noisy configuration leaves many atoms unassigned because thermal disorder and strong elastic distortion obscure the local coordination environment. After denoising using the log-probability model, the local atomic arrangement becomes substantially more coherent and PTM no longer produces large ``unknown'' regions, indicating that the learned score field removes high-frequency structural noise while preserving the underlying microstructure.

The resulting phase map obtained from the log-probability model is shown in Figure~\ref{fig:babak structure-2}b. The model identifies spatially coherent HCP-like and $\omega$-like domains throughout the shocked sample and identifies the spatially heterogeneous nature of the transformation. To quantify classification confidence, Figure~\ref{fig:babak structure-2}c shows the top-2 logit margin, $l_{(1)} - l_{(2)}$, where $l_{(1)}$ and $l_{(2)}$ are the largest and second-largest logits for each atom, respectively. Large margins correspond to atoms that strongly match a single structural prototype, whereas smaller margins occur near interfaces and highly distorted regions where local environments are structurally ambiguous.

Figures~\ref{fig:babak structure-2}d,e characterize the temporal evolution of the HCP$\rightarrow\omega$ transformation during shock propagation. Figure~\ref{fig:babak structure-2}d shows progressive nucleation and growth of $\omega$ domains between frames 10 and 35. To obtain a continuous structural descriptor, we define a differentiable order parameter using the logit difference $l_{\omega} - l_{\mathrm{HCP}}$. Although the values remain predominantly negative because HCP-like environments still dominate the system, the distribution systematically shifts toward larger values as the simulation evolves (Figure~\ref{fig:babak structure-2}e). This shift reflects the increasing prevalence of $\omega$-like local environments during shock loading and provides a continuous description of the phase transformation beyond discrete template assignment. Additional comparisons with PTM-based classification and the chemistry-agnostic model are provided in the Supporting Information (Figure~\ref{fig:shock_si_compare}). The chemistry-agnostic model produces qualitatively similar phase maps, indicating that the structural recognition is not solely driven by chemical identity. In contrast, PTM frequently maps highly distorted $\omega$-like environments onto HCP, BCC, or ``unknown'' categories under strong strain and disorder.

\section*{Discussion}

Despite being trained exclusively on ordered crystalline structures mapped to AFLOW prototypes and augmented only with synthetic elastic and thermal perturbations, the log-probability model remains effective across several extrapolation regimes, including thermal distortions, point defects, mixed solid--liquid interfaces, and shock-induced phase coexistence. By learning a global scalar log-density $\log \hat{P}_\theta(\boldsymbol{r})$ whose gradient defines the denoising direction, the model unifies three tasks that are typically treated separately: denoising perturbed configurations, assigning crystal phase labels, and providing continuous, physically interpretable OPs derived from the same per-phase logit landscape $l_{ac}$. Despite its relatively simple training setup, the model provides a reusable, phase-agnostic representation that transfers across diverse crystal prototypes and downstream tasks.

This unified view leads to practical advantages over conventional symmetry-based and template-based approaches such as CNA or PTM. In noisy MD trajectories and high-temperature (e.g.\ DC3) snapshots at or above the nominal melting point, where hard geometric thresholds often fail,  logP maintains high classification accuracy and can recover the correct prototype within a few denoising steps. Per-atom logits $l_{ac}$ admit a simple and physically motivated interpretation that is approximately related to the squared distance with respect to the ideal structure. They act as smooth OPs that track gradual structural transformations, as illustrated by the Bain and Burgers paths and by the spatial variation across a water--ice interface. In shock-compressed Ti, the modelidentifies coexisting HCP-like, BCC-like, and $\omega$-like regions under strong strain and disorder, while template-based PTM, which lacks an explicit $\omega$ prototype, necessarily maps structurally $\omega$ regions onto HCP, BCC, or ``unknown'' labels. The resulting OP fields $l_{ac}$ and phase maps ($\arg\max_c l_{ac}$ with adjustable thresholds) provide a detailed description of phase coexistence and interfaces as well as physically interpretable OPs of similarity to structural prototypes, offering a natural starting point for quantitative analysis of $\omega$-phase nucleation and growth mechanisms in dynamically loaded Ti.

Our work is also closely related to earlier deep-learning approaches for crystal-structure classification, most notably the diffraction-image classifier of Ziletti \textit{et al.}~\cite{Ziletti2018NC}. That study demonstrated that convolutional neural networks operating on 2D diffraction fingerprints can achieve nearly perfect classification of a small set of elemental crystal families and can remain robust under substantial disorder and defects. However, the classifier operates on reciprocal-space images and produces global class probabilities for a limited number of prototype classes. In contrast, the present log-probability model works directly on real-space atomic graphs, scales to hundreds of AFLOW prototypes and thousands of elemental and binary structures, and outputs per-atom, per-phase $l$ values whose gradients define denoising displacements. This allows robust classification, denoising, and OP extraction to be handled within a single model, with spatial resolution sufficient to analyze interfaces, defects, and complex microstructures far beyond the scope of purely image-based classifiers.

A distinctive feature of the present approach is that it makes confidence and ambiguity in phase assignments directly visible. Regions that closely resemble a given prototype have large, sharply peaked $l$ for that class, whereas atoms near phase boundaries, defect cores, or strongly distorted environments exhibit reduced maxima or competing phase preferences. While this does not constitute a formal statistical uncertainty estimate in the sense of Bayesian or ensemble methods, it provides an intuitive, data-driven measure of how well each local environment matches the available prototypes. In practice, this graded view helps distinguish bulk-like regions from structurally atypical ones and complements hard categorical labels produced by existing tools.

Our current implementation has a practical computational limitation. The model is built on the full MACE architecture used in MACE-MP, with relatively wide hidden representations and multiple interaction layers. Therefore computational costs are essentially the same as MACE-MP. While this choice is advantageous for accuracy and transferability, it also makes the model memory intensive. For very large atomistic configurations (e.g., shock simulations or large-scale MD snapshots with $> 10^5$ atoms), naive evaluation of the full model on a single GPU can lead to out-of-memory failures. In practice, this can be mitigated by domain decomposition techniques used in MLIP inference.
Another mitigation strategy is half-precision inference for large structures without discernible discrepancy compared to single or double-precision evaluations in our tests. Another limitation is that it considers so many competing phases that its classification accuracy may be lower with zero or few denoising steps for ``tricky'' phases such as close-packed structures with different long-range stacking patterns. This can be mitigated with more denoising steps, or by focusing on outputs of a smaller pool of candidate structures. It is also possible that the frozen featurization layers of MACE-MP were relatively insensitive to subtle difference in long-range ordering, and therefore should be fine-tuned for improved classification accuracy.

\section*{Conclusion}

Overall, this work establishes a unified probabilistic framework for analyzing noisy atomic configurations. The log-probability model does not only denoise structures; it provides a probabilistic framework that simultaneously explains, classifies, and quantifies structural order in crystalline materials, and that remains robust under challenging extrapolation regimes such as high-temperature DC3 structures at or above the melting point and shock-compressed Ti. As our logP method relies on no prior knowledge of specific crystalline phases other than the ideal structure, it can in principle be extended to additional prototypes and chemistries once the reference structures are included in training.
Our probabilistic OPs, distinguished by their ease of development, broad applicability and physics distance-like interpretation near known prototypes, will facilitate novel investigations in the modeling of phase transformations. Extending this framework to jointly model crystalline, liquid, and amorphous phases, to incorporate chemically disordered alloys, and to couple log-probability learning with generative sampling or automated prototype discovery are promising directions for future work. Such developments would further strengthen the role of log-probability models as general tools for automated structure analysis, phase mapping, and data-driven thermodynamics in computational materials science.



\section*{Data availability}
The curated datasets, representative molecular dynamics trajectories, inference outputs, and scripts used in this work are publicly available at Zenodo: \url{https://doi.org/10.5281/zenodo.20089143}.

\section*{Code availability}
The full code is available in the NPS repository at \href{https://github.com/kha8128/NPS/tree/logp-model/NPS/logp}{https://github.com/kha8128/NPS/tree/logp-model/NPS/logp}.


\bibliographystyle{naturemag}
\bibliography{ref}

\section*{Acknowledgment}
This work was performed under the auspices of the U.S.\ Department of Energy by Lawrence Livermore National Laboratory under Contract DE-AC52-07NA27344. This work was funded by the Laboratory Directed Research and Development (LDRD) Program at LLNL under project tracking code 25-ERD-002 (HK, JK, FZ), 23-SI-006 (SH, VL). Computing support for this work came from the Lawrence Livermore National Laboratory Institutional Computing Grand Challenge program.

\section*{Author contributions}
FZ developed and implemented the model and led the research. HK performed the computational experiments and data analysis. BS, SH, VL and JK contributed technical advice and data. JK and VL secured funding. HK and FZ wrote the paper with input from all authors.

\section*{Competing interests}
The authors declare no competing interests.
    

\newpage
\appendix
\renewcommand\thefigure{S.\arabic{figure}}
\renewcommand\thetable{S.\arabic{table}}

\setcounter{figure}{0}
\setcounter{table}{0}
\section*{Supplemental material}

\section{Characterization of extrapolation regimes}

To clarify whether each evaluation setting probes interpolation or extrapolation beyond the training perturbation regime, we summarize the relationship between the training data and all test cases in Table~\ref{table:training_test_ood}. We distinguish between coverage at the level of structural prototypes, chemical composition, and perturbation regime. While most test cases involve prototypes and compositions present in the training set, several scenarios probe extrapolation in physically meaningful directions, including noise amplitudes beyond the training range, high-temperature disorder and defects, liquid phases not seen during training, and large non-equilibrium strains. This breakdown makes explicit which aspects of the evaluation correspond to interpolation and which probe physically meaningful extrapolation regimes.

\begin{table}[h]
\centering
\caption{Summary of training/test overlap and extrapolation regimes. “Proto.” indicates structural prototype coverage, “Comp.” composition coverage, and “Pert.” whether perturbations exceed the training regime.}
\begin{tabular}{lllll}
\hline
Case & Proto. & Comp. & Pert. & Extrapolation type \\
\hline
MP test set & Yes & Yes & No & interpolation \\
Ag A\_hP2\_194 & Yes & Yes & Yes & noise (high $\sigma$) \\
DC3 & Yes & Yes & Yes & thermal / defects \\
water--ice & Ice only & Yes & Yes & phase (liquid) \\
shock Ti & Yes & Yes & Yes & strain (non-eq.) \\
\hline
\end{tabular}
\label{table:training_test_ood}
\end{table}

\section{Effect of pretrained model featurization on optimization.}

To quantify the benefit of model pretraining, we compared two training protocols on the curated crystal-structure dataset. In the \textbf{pretrained} setup, we initialize the model from MACE-MP weights and optimize the log-probability objective on top of the pretrained equivariant representations. In this setup, the model was initialized from pretrained MACE-MP weights and subsequently optimized under the same log-probability objective. In the \textbf{from-scratch} setup, all parameters of the MACE architecture are randomly initialized and trained jointly under the same objective.

To more rigorously assess the impact of pretraining, we compared final model performance after full convergence using identical architectures and training settings. Both pretrained and randomly initialized models achieve essentially identical classification accuracy (99.83\%). However, the pretrained model yields a slightly lower RMSE (0.0015~\AA\ vs 0.0019~\AA) and improved score-matching loss, while the from-scratch model attains a lower classification loss. These results indicate that pretraining primarily benefits the geometric (score-based) component of the objective, consistent with the energy/force training of MACE-MP, but does not substantially improve classification performance or overall optimization of the full log-probability objective. Overall, initialization from pretrained MACE-MP weights provides only modest improvements in denoising performance and optimization stability.

\begin{figure}[h!]
    \centering
    \includegraphics[width=0.7\linewidth]{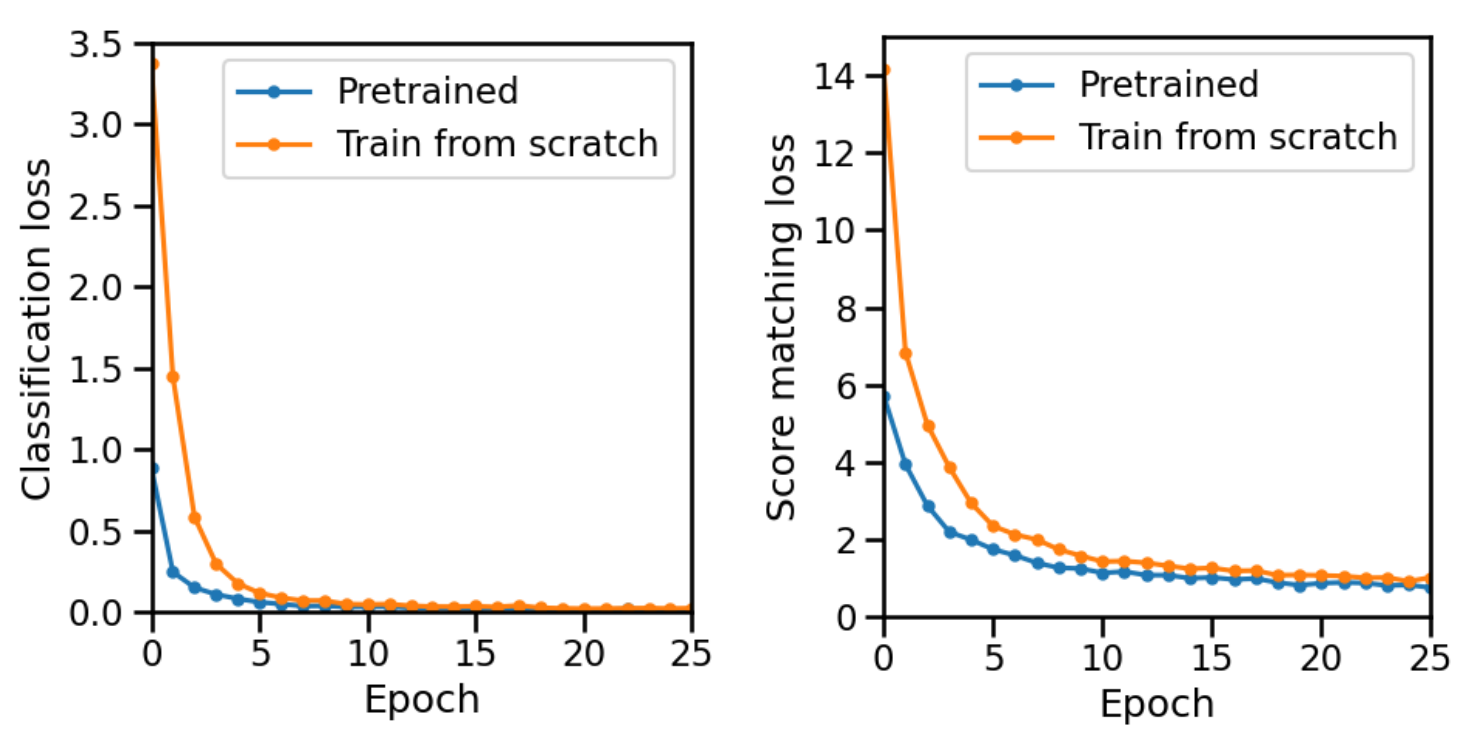}
    \caption{Effect of pretrained model initialization on optimization. Models initialized from pretrained MACE-MP weights (blue) begin from lower initial losses and exhibit slightly improved score-matching behavior compared to models trained from scratch (orange). However, both approaches ultimately converge to similar classification performance, indicating that pretraining provides only modest benefits for optimization of the full log-probability objective.}
\label{fig:pretrain_scratch_comparison_supp}
\end{figure}

\section{Machine-learning model details}
The MACE-MP checkpoint \texttt{2024-01-07-mace-128-L2\_epoch-199.pt} was used to initialize the \textbf{pretrained} training. This checkpoint is a \texttt{ScaleShiftMACE} model with cutoff radius $r_\mathrm{max}=6.0$~\AA, 10 radial Bessel basis functions, a polynomial with cutoff $p=5$, 2 interaction layers, and 128 scalar channels in the initial node embedding. The model uses spherical harmonics up to $l=3$, as indicated by tensor products involving $1\times 0e + 1\times 1o + 1\times 2e + 1\times 3o$, and includes 89 atomic-number embeddings. The first interaction maps $128\times 0e$ features to $128\times 0e + 128\times 1o + 128\times 2e + 128\times 3o$, while the product basis reduces the propagated features to $128\times 0e + 128\times 1o + 128\times 2e$. The second interaction layer uses the same hidden-channel scale and is followed by the standard MACE linear and nonlinear readout blocks. 

For the log-probability model, the equivariant MACE representation is coupled to a prototype-resolved decoder that predicts per-atom logits $l_{ac}$ for each structural class $c$. The global log-density is then constructed from these logits through the log-sum-exp aggregation described in the Methods. In the pretrained setup, the model parameters were initialized from the MACE-MP checkpoint, while the newly introduced log-probability decoder parameters were initialized randomly. The entire model, including both the pretrained MACE layers and the newly added decoder, was then jointly optimized under the combined score-matching and classification objective. In the from-scratch setup, all parameters in the same architecture were randomly initialized and trained under the same objective.

In the chemistry-agnostic implementation, all atoms were reassigned to a single shared pseudo-species (H) prior to graph construction and inference. As a result, the model could not use chemical identity information and instead relied entirely on local geometric and topological environments for phase discrimination. All other architectural components, graph construction procedures, cutoff radii, and training objectives were kept identical to the chemistry-aware model.

\section{Comparison against CHILL+ at a water--ice interface}
To compare the probabilistic log-probability framework against a conventional local structural classifier, we evaluated the OVITO CHILL+ algorithm on the same water--ice coexistence trajectory used in the main text. CHILL+ classifications were computed for the original thermally perturbed configuration ($k=0$) and for denoised configurations after $k=2$ and $k=8$ denoising iterations using a cutoff radius of 2.1~\AA, which produced the most stable classifications among the tested parameter choices.

\begin{figure}[h!]
    \centering
    \includegraphics[width=\linewidth]{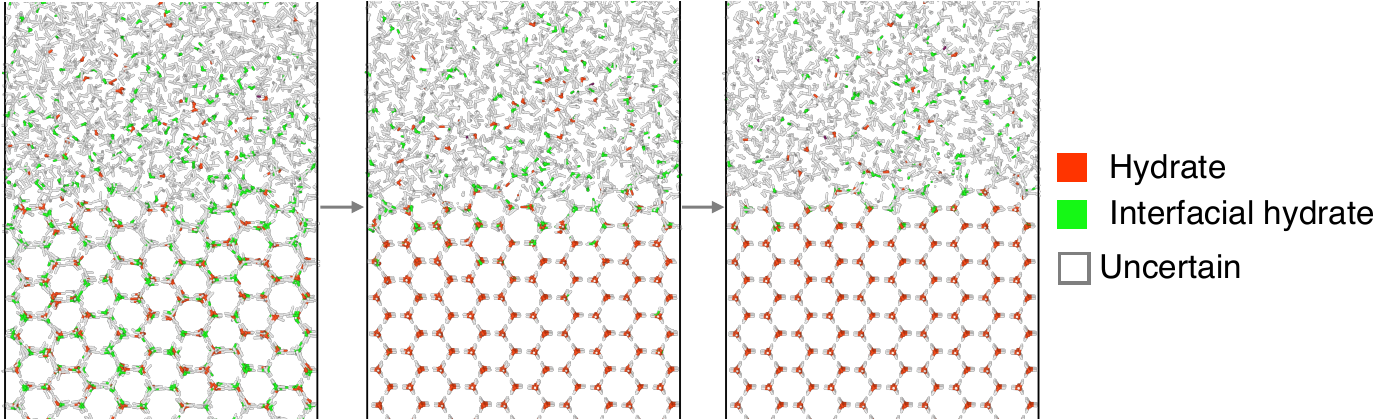}
    \caption{CHILL+ classification of the water--ice coexistence trajectory at denoising steps $k=0$, $2$, and $8$ using a cutoff radius of 2.1~\AA. Although denoising improves recognition of the crystalline ice slab, CHILL+ continues to produce fragmented hydrate- and interfacial-hydrate classifications near the solid--liquid boundary and within thermally fluctuating liquid environments.}
\label{fig:water_chill_compare}
\end{figure}

Figure~\ref{fig:water_chill_compare} shows that CHILL+ identifies the bulk crystalline region as predominantly hydrate-like after denoising, but the interfacial region remains fragmented and spatially heterogeneous even at late denoising stages. In particular, CHILL+ continues to assign mixtures of hydrate and interfacial-hydrate labels near the solid--liquid boundary, with isolated ordered motifs also appearing within the liquid region due to transient tetrahedral fluctuations.

In contrast, the log-probability framework presented in the main text yields a smoother and more spatially coherent probabilistic transition between crystalline and liquid environments. Rather than relying on discrete local template assignments, the probabilistic OP continuously tracks the degree of ice-like order and naturally accommodates partially ordered interfacial environments without introducing fragmented classifications.

\section{Additional results on thermally perturbed DC3 structures}
\begin{figure}
    \centering
    \includegraphics[width=\linewidth]{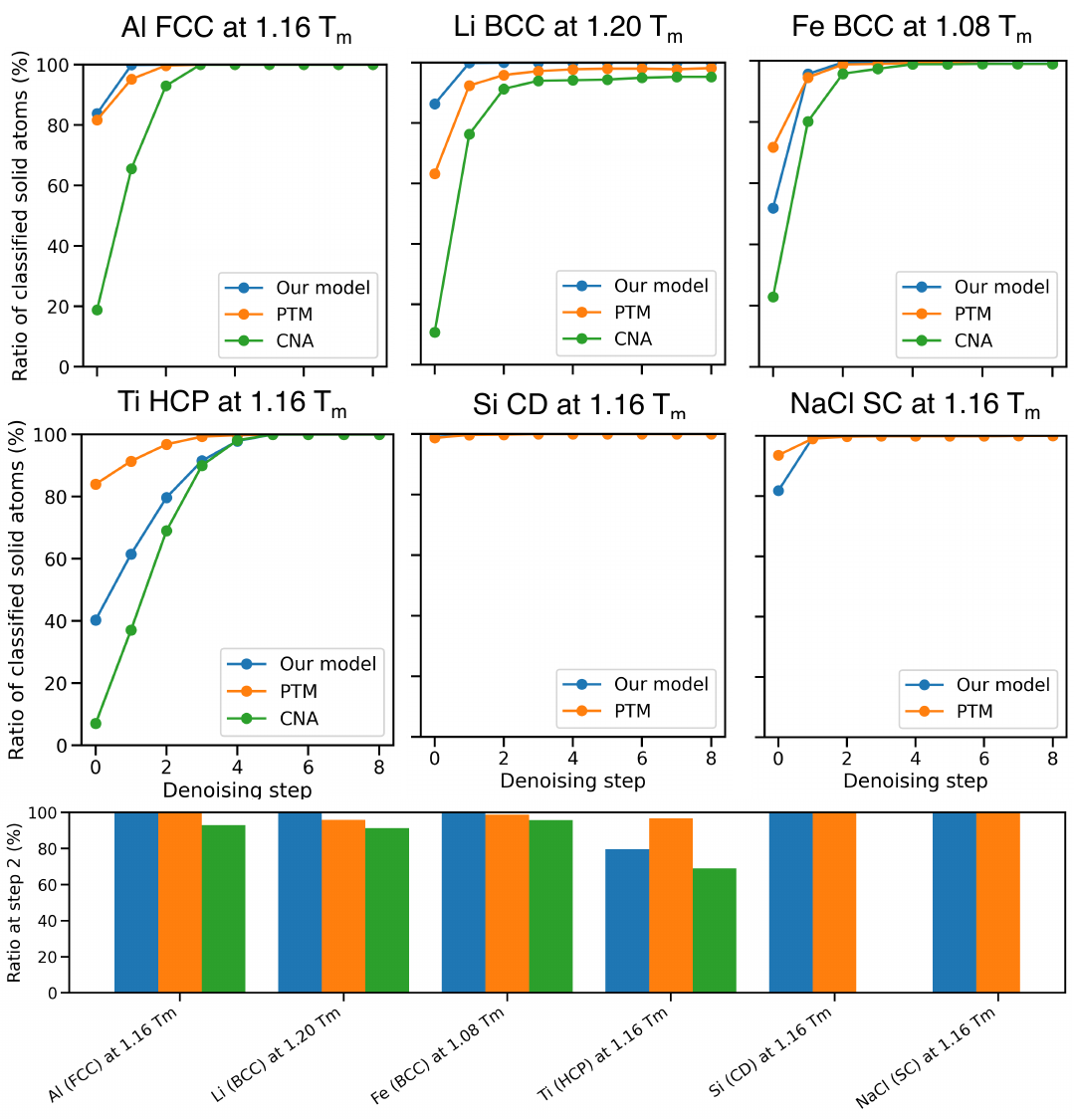}
    \caption{Classification accuracy on thermally perturbed structures from the DC3 database. Top panels: accuracy versus denoising step for representative elemental and binary systems at high temperatures above their melting points, comparing the log-probability model to PTM and CNA. The model reaches higher accuracy within fewer denoising iterations and often achieves 100\% accuracy by step~3. Bottom panel: summary of classification accuracy at step~2 across all tested systems, showing consistently superior early-stage performance over PTM and CNA under strong thermal disorder. For each denoising step $k$, PTM and CNA are evaluated on the same coordinates as the log-probability model, i.e.\ on the configuration obtained after $k$ log-probability denoising steps (with $k=0$ corresponding to the original DC3 snapshot).}
    \label{fig:DC3}
\end{figure}

The DC3 database provides high-temperature MD snapshots that combine strong vibrational disorder with occasional non-thermal defects (e.g., vacancies/interstitials), posing a stringent test for local, template-based structure identifiers. This comparison is therefore stricter than applying PTM/CNA only to the raw MD snapshots: PTM and CNA are also given the benefit of the same log-probability denoised coordinates at each step. The remaining performance gap therefore reflects differences in the structural assignment rule rather than simply the availability of denoising. Fig.~\ref{fig:DC3} reports classification accuracy as a function of denoising step for representative elemental and binary systems. For each system, we start from the highest-temperature snapshot available ($k=0$) and apply $k=1,\dots,8$ log-probability denoising steps. To enable a like-for-like  comparison, PTM and CNA are evaluated on the same coordinates at each step  $k$ (i.e., on the configuration produced  after $k$ log-probability denoising steps), so differences reflect the  classifiers rather than differences in denoising. Overall, the model reaches high accuracy with fewer denoising iterations, while template-based methods can plateau when defect-containing local environments remain difficult to match to ideal templates.

\section{Effect of elastic-strain augmentation on prototype classification}

\begin{figure}[h!]
    \centering
    \includegraphics[width=\linewidth]{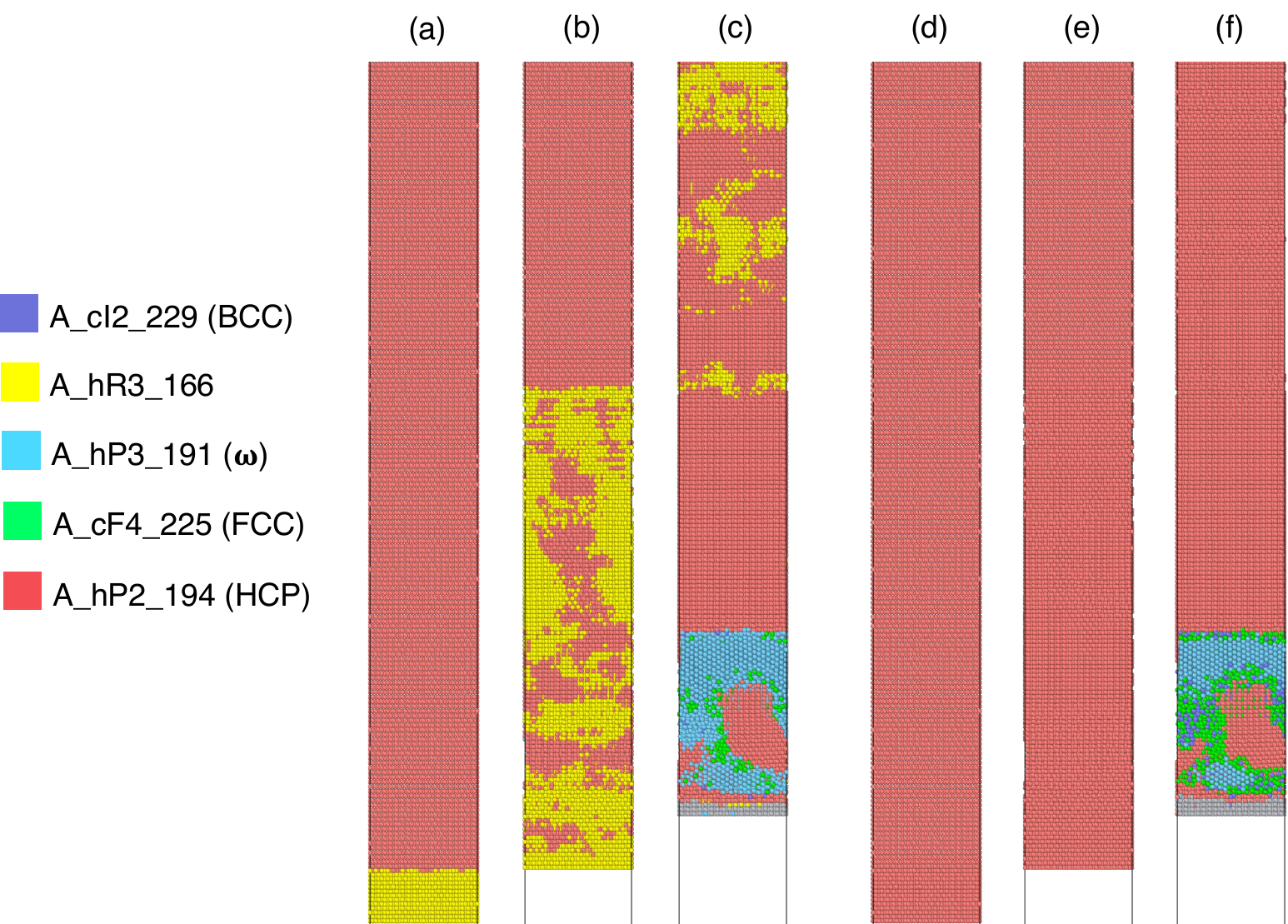}
    \caption{Effect of elastic-strain augmentation on prototype classification in shocked HCP Ti. (a–c) Frames 0, 7, and 14 from a uniaxial shock-compression trajectory of HCP Ti (shock applied from the bottom), using a model retrained without random elastic-strain augmentation. The uniaxially compressed region is systematically misclassified as the rhombohedral A\_hR3\_166 prototype (space group 166, yellow), indicating that the network explains strain-induced distortions by switching prototypes rather than recognizing them as deformed HCP. (d–f) The same frames from the same trajectory evaluated with a model trained with elastic-strain augmentation, which correctly classifies the entire shocked region as HCP (red) and eliminates the spurious A\_hR3\_166 pocket, demonstrating that elastic-strain augmentation is essential for robust generalization to strongly compressed microstructures.
}
\label{fig:with_strain_comparison}
\end{figure}

To assess the role of elastic-strain augmentation, we retrained the $\log \hat{P}_\theta$ model with the same architecture and hyperparameters but without the random elastic-strain transformations applied during pretraining and fine-tuning. We then evaluated both models on a uniaxial shock-compression trajectory of HCP Ti (the same trajectory as in the main text). As shown in Fig.~\ref{fig:with_strain_comparison}, the model trained without elastic-strain augmentation systematically misclassifies the uniaxially compressed HCP region as the rhombohedral \texttt{A\_hR3\_166} prototype (space group 166), effectively explaining the strain-induced distortions by switching to a different prototype rather than recognizing them as elastically deformed HCP. In contrast, the model trained with elastic-strain augmentation correctly preserves the HCP label throughout the shocked region for all frames. This ablation confirms that elastic-strain augmentation is critical for robust generalization to strongly compressed microstructures and suppresses spurious prototype switching under large uniaxial strain.

\section{Additional shock-compressed Ti comparisons}
To further assess robustness under strong nonequilibrium deformation, we compare the full log-probability model against both PTM and a chemistry-agnostic variant of the model on the shock-compressed Ti trajectory discussed in the main text.

\begin{figure}[h!]
    \centering
    \includegraphics[width=\linewidth]{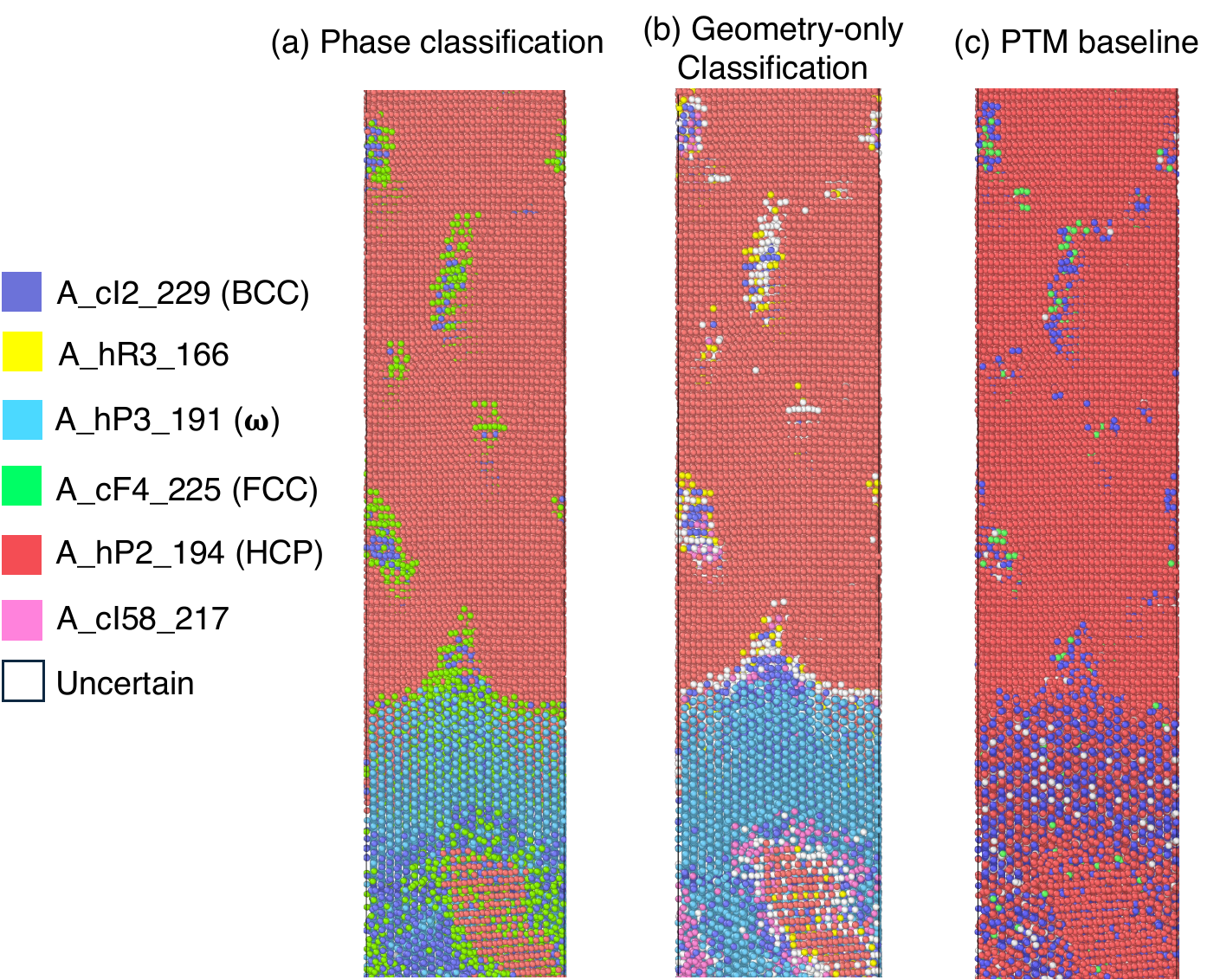}
    \caption{Comparison of structural classification methods on shock-compressed Ti (frame 75). (a) Full log-probability model. (b) Chemistry-agnostic model. (c) PTM classification. The chemistry-agnostic model produces phase maps qualitatively similar to the full model, whereas PTM leaves many distorted regions unassigned or maps $\omega$-like environments onto alternative templates.}
\label{fig:shock_si_compare}
\end{figure}

Figure~\ref{fig:shock_si_compare} compares the resulting phase maps for frame 75. The chemistry-agnostic model produces qualitatively similar HCP/$\omega$ domain structures, indicating that the classification is primarily driven by local geometry rather than explicit chemical identity. PTM, in contrast, frequently labels highly distorted regions as ``unknown'' or maps $\omega$-like environments onto HCP or BCC templates because an explicit $\omega$ prototype is not included in the PTM template set.

These comparisons support the interpretation that the learned log-probability field provides more robust structural recognition under strong strain and disorder than discrete template-based approaches.

\end{document}